\documentclass[final,5p,times]{elsarticle}

\usepackage{graphicx}
\usepackage{amsmath}
\usepackage{amssymb}
\usepackage{hyperref}
\usepackage{lineno}
\usepackage{subcaption}
\usepackage{multirow}
\usepackage{dsfont}
\usepackage{float}  

\usepackage{xspace}

\usepackage[superscript]{cite}

\usepackage{booktabs} 
\modulolinenumbers[5]

\journal{The Lancet Digital Health}

\begin{document}

\begin{frontmatter}


\title{Observing the unobserved confounding through its effects:\\
toward randomized trial-like estimates from real-world survival data}

\author{
Vasiliki Stoumpou$^{1}$,
Dimitris Bertsimas$^{1,2}$,
Samuel Singer$^{3}$,
Georgios Antonios Margonis$^{2,3,4}$\\[1em]
\small $^{1}$Operations Research Center, Massachusetts Institute of Technology, Cambridge, MA, USA\\
\small $^{2}$Sloan School of Management, Massachusetts Institute of Technology, Cambridge, MA, USA\\
\small $^{3}$Department of Surgery, Memorial Sloan Kettering Cancer Center, New York, NY, USA\\ 
\small $^{4}$Charité-Universitätsmedizin Berlin, Corporate Member of Freie Universität Berlin and Humboldt-Universität zu Berlin, Berlin, Germany
\texttt{margonig@mskcc.org}
}

\begin{abstract}

\indent 

Background:
Randomized controlled trials (RCTs) are costly, time-consuming and often infeasible, whereas treatment-effect estimation from observational data is limited by unobserved confounding.

Methods:
We developed a three-step framework to address unobserved confounding in observational survival data. First, we infer a latent prognostic factor ($\tilde{U}$) from restricted mean survival time (RMST) discrepancies between patients with similar observed prognostic factors, the same treatment assignment, and divergent outcomes, based on the premise that the aggregate effect of unmeasured prognostic factors can be inferred even when individual factors cannot be observed.  Second, we balance $\tilde{U}$ together with observed baseline prognostic factors between treatment groups using prognostic matching, entropy balancing, or inverse probability of treatment weighting. Third, we apply multivariable survival analysis to estimate hazard ratios (HRs). We evaluated the framework in three settings: three observational cohorts with RCT benchmark HRs, two RCT cohorts, and six multicenter observational cohorts.


Results: 
In three observational cohorts (nine cohort-method comparisons), balancing $\tilde{U}$ improved agreement with benchmark trial HRs in all comparisons; in the strongest settings, it reduced absolute log-HR error by approximately ten-fold relative to adjustment using observed covariates alone (mean reduction, 0.344; p=0.001). In the two RCT cohorts, $\tilde{U}$ was balanced across treatment arms (most SMDs <0.1) and adjusting for it had little effect on estimated treatment log-HRs (mean absolute change $\simeq$ 0.08).  Across six multicenter cohorts, balancing $\tilde{U}$ reduced cross-center dispersion in chemotherapy log-HR estimates for overall- and recurrence-free survival (mean reduction 0.147; $p=0.016$) when balancing was performed within centers to address residual confounding, and narrowed cross-center survival differences in 75\% to 100\% of informative comparisons when populations were directly balanced across centers to account for differences in patient case mix. 

Conclusions:
Inferring and balancing a latent prognostic signal may reduce unobserved confounding in observational survival analyses and support more reliable treatment-effect estimation from real-world data.

\end{abstract}


\end{frontmatter}

\section{Introduction}

Observational real-world data are increasingly used to evaluate treatment effects, particularly when randomized controlled trials (RCTs) are infeasible, underpowered, outdated, or not fully representative of contemporary practice. However, such analyses remain vulnerable to bias, including immortal time bias, selection bias, and confounding by indication. Although recent methodological advances have improved our ability to address some of these problems, and standard regression, matching, and weighting approaches can mitigate confounding from measured covariates, unobserved confounding remains fundamentally difficult to resolve. By definition, it is not possible to adjust for variables that are not measured; one cannot regress or match on what is not observed. 

\indent As a result, even after careful adjustment for measured covariates, treatment groups may still differ in important unmeasured aspects of prognosis, potentially leading observational studies to produce biased or even misleading treatment-effect estimates. This challenge is especially relevant in surgical oncology, where decisions regarding surgery and perioperative treatment often depend on nuanced aspects of disease severity, patient fitness, and clinician judgment that are not fully captured in routine datasets. In practice, this problem is often addressed not by solving it, but by assuming that unobserved confounding is negligible for the clinical question at hand. Yet that assumption is frequently implausible, as illustrated by settings in which treatment-effect estimates from observational data were later contradicted by randomized evidence \cite{nussbaum2016preoperative, bonvalot2020preoperative}. These discrepancies highlight the need for methods that can account, at least indirectly, for latent prognostic differences that remain outside the recorded data.\\

\begin{center}
\fbox{
\begin{minipage}{0.46\textwidth}
\section*{Research in context}

\noindent\textbf{Evidence before this study}

Unobserved confounding arising from unknown, unmeasured, or unrecorded prognostic factors remains a fundamental challenge in clinical outcomes research. When such factors are imbalanced across treatment groups, treatment effect estimates can be biased even after adjustment for observed covariates. Existing approaches generally do not directly address this problem; instead, they rely on the assumption that unobserved confounding is limited or negligible in the clinical setting under study. As a result, the validity of many observational analyses depends heavily on design restrictions and unverifiable assumptions.

\noindent\textbf{Added value of this study}
Existing methods for treatment-effect estimation in real-world data, including inverse probability of treatment weighting, entropy balancing, and prognostic matching, adjust only for observed covariates and therefore remain fundamentally limited by unobserved confounding. The added value of this study is that it extends these standard frameworks by incorporating a patient-level latent quantity, $\tilde{U}$, that indirectly captures the aggregate effects of unobserved prognostic factors. Thus, rather than only balancing what is measured, the proposed approach seeks to also balance part of what is unmeasured. Empirically, this added value was supported by improved agreement with benchmark randomized trials, by preservation of balance of $\tilde{U}$ across randomized trial arms, and by findings in multicenter colorectal liver metastasis cohorts showing that $\tilde{U}$ captured both hidden factors contributing to treatment-selection bias within cohorts and hidden differences in case mix across centers. These are sources of heterogeneity that conventional observed-covariate methods cannot directly address.

\noindent\textbf{Implications of all the available evidence}
This framework provides a practical strategy to strengthen causal inference from real-world data in the presence of likely unobserved confounding. By augmenting standard adjustment methods with a patient-level estimate of latent prognostic burden, it may help observational analyses better approximate trial-like treatment-effect estimates. In doing so, it could be particularly useful for revisiting unresolved clinical debates in settings where randomized evidence is unavailable, limited, outdated, or difficult to generalize to contemporary practice.
\end{minipage}
}
\end{center}

\indent In this study, we propose a framework to address this problem by inferring and balancing a latent prognostic signal that reflects the aggregate effect of otherwise unobserved prognostic factors. Rather than attempting to recover each missing factor individually, the framework aims to capture their combined prognostic influence before estimating treatment effects as hazard ratios (HRs). We evaluate the framework in three complementary settings. First, in three observational surgical oncology cohorts, we test whether it moves treatment-effect estimates closer to corresponding RCT benchmarks. Second, in two oncology RCT datasets, where confounding should be absent by design, we test whether it leaves the trial-reported HRs largely unchanged. Third, in multicenter colorectal liver metastasis (CRLM) data, we test whether the framework reduces two distinct forms of between-center heterogeneity: hidden bias in who receives adjuvant chemotherapy within a center, reflected in variation in chemotherapy HR estimates across centers, and hidden differences in patient case mix across center populations, reflected in residual cross-center survival differences after balancing recorded covariates.

\section{Methods}

We first provide an overview of the proposed methodology and then describe each component in detail. Our approach aims to transform an observational cohort to a population that closely mimics the conditions of an RCT, namely balance in both observed and unobserved covariates. The ultimate goal is to enable reliable treatment effect estimation within a survival analysis framework, which can approximate the results we would expect if the data had come from an actual RCT evaluating the same therapy.

The first step addresses unobserved prognostic factors and constitutes the main contribution of our methodology. We estimate an unobserved feature $\tilde{U}$, which serves as a latent prognostic factor capturing outcome-related differences not explained by the observed covariates. In other words, this variable is constructed to capture variation in the outcome that cannot be adequately justified by the observed features alone.

The second step incorporates this latent factor together with the observed covariates in the balancing procedure, using standard balancing methods such as prognostic matching followed by sample weighting \cite{bertsimas2024road}, entropy balancing \cite{hainmueller2012entropy}, or inverse propensity weighting \cite{rosenbaum1987model}. By balancing both observed covariates and the inferred latent prognostic factor across treatment arms, this step aims to address both observed and unobserved confounding.

The third step involves multivariable analysis, where we fit a Cox proportional hazards model, which serves as a final correction for any residual observed confounding and remaining imbalances. The coefficient of the treatment variable in the resulting model is then utilized to estimate the treatment Hazard Ratio, which is used to evaluate the treatment effect.

\begin{figure*}[htbp!]
    \centering
    \includegraphics[width=0.7\linewidth]{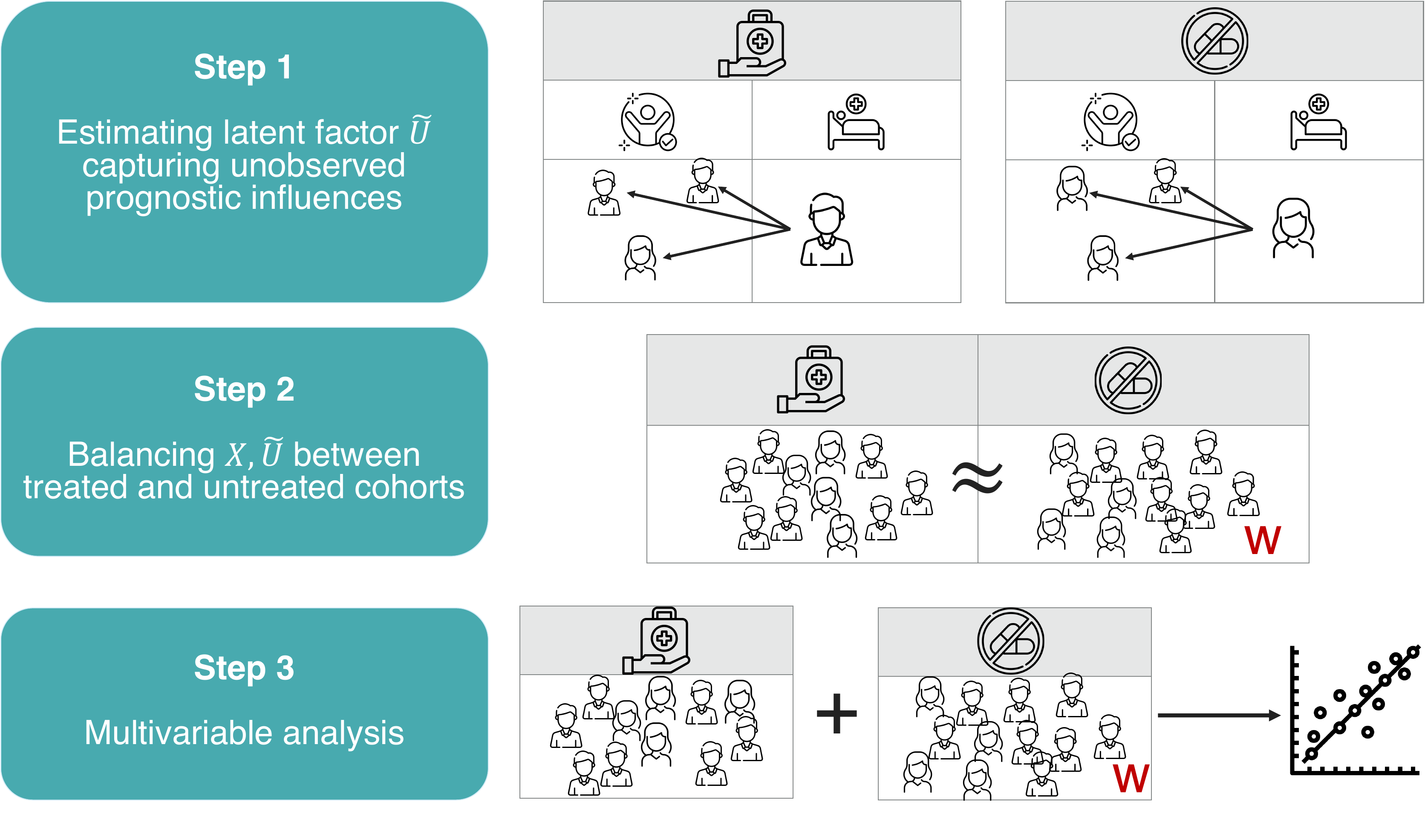}
    \caption{Full pipeline of the proposed framework.}
    \label{fig:full_pipeline}
\end{figure*}

The full pipeline is depicted in Figure \ref{fig:full_pipeline}. The following Sections elaborate on each of these three steps.

\subsection{Step 1: Constructing the latent factor \texorpdfstring{$U$}{U}}\label{sec:latent_factor}

This step introduces our core methodological contribution: estimating a hidden prognostic factor that can later be used to correct for unobserved confounding.

Our goal is to define a scalar quantity $\tilde{U}_i$ that captures the portion of patient $i$'s outcome that cannot be explained by observed covariates. One way to approach the construction of $\tilde{U}_i$ is to utilize patient $i$'s nearest neighbors, namely patients with similar observed characteristics, treatment, and clinical presentations. Such patients are expected to have similar outcomes to patient $i$; however, in practice, individuals with nearly identical observed profiles may experience very different survival trajectories (e.g., recurrence vs.\ non-recurrence), suggesting the influence of unobserved patient-level factors.

Intuitively, we quantify how ``unexpected'' a patient's outcome is relative to similar patients within the same treatment arm. When two such patients exhibit different outcomes, the discrepancy primarily reflects latent factors; as these patients are similar in their observed features and share the same treatment assignment, the difference in outcomes is mostly driven by unobserved influences such as biological differences, undocumented comorbidities, or other unmeasured sources of heterogeneity. Quantifying this discrepancy between the neighbors therefore provides a way to capture the influence of these unmeasured factors on the outcome: a small discrepancy suggests that unobserved factors have limited impact, whereas a larger discrepancy indicates a stronger latent influence. Importantly, we do not attempt to recover any specific latent variable, as this would be infeasible for factors that are not directly observed or even known. Instead, $\tilde{U}$ should be interpreted as a summary measure of the combined influence of unobserved factors affecting the outcome, which is expected to be balanced in randomized controlled trials.

There are two important nuances in the construction of $\tilde{U}_i$, without which differences in outcomes between neighboring patients cannot be safely attributed to unobserved factors.

The first is that focusing on neighbors within the same treatment arm is crucial. Under this restriction, we can reasonably assume that the interaction between treatment and the observed covariates is approximately the same for nearby patients in the same treatment arm, and that the baseline effect of observed covariates is similar for nearby patients in the untreated arm. Consequently, differences in outcomes between such neighbors are not attributed to treatment-related heterogeneity, but to hidden, underlying factors that affect their outcomes in different ways.

The second is that $\tilde{U}_i$ must satisfy a directional sign convention. By forcing the comparison to occur against neighbors with the opposite survival trajectory, we ensure that the sign of the residual reflects latent influence with respect to the observed outcome. Under this convention, negative values correspond to latent adverse factors associated with worse-than-expected survival, whereas positive values correspond to latent favorable factors associated with better-than-expected survival, conditional on the observed covariates and treatment status. For example, among patients who do not receive chemotherapy despite having similar observed disease characteristics and apparent indication for treatment, negative values may reflect unmeasured frailty or comorbidity burden that both discouraged treatment and contributed to earlier death, whereas positive values may reflect preserved physiologic reserve or other favorable unmeasured characteristics in patients who remained untreated for non-clinical reasons, such as treatment refusal. The farther $\tilde{U}_i$ lies from zero, the stronger the inferred influence of these unobserved factors. Without this directional selection of neighborhoods, a patient with an unexpectedly early event could have a value of $\tilde{U}_i$ similar to that of a patient with unexpectedly long survival, as contributions from neighbors with mixed outcome trajectories could cancel out.

These two elements are essential: selecting neighbors purely based on proximity in the covariate space would remove the directional interpretation of the residual, while selecting neighbors from a different treatment arm would reintroduce treatment effects into the comparison. In either case, the interpretation of the residual as reflecting latent factors would be undermined.

\subsubsection{Directional neighbor sets (within arm)}

We now formalize the neighborhood-based construction introduced above. Let $X_i$ denote the observed covariates of patient $i$, $T_i$ their observed follow-up time, and $E_i \in \{0,1\}$ the event indicator ($1$ = event observed; $0$ = censored). Let $V_i$ denote unobserved factors (e.g., frailty, latent tumor aggressiveness, undocumented comorbidities) that may influence both treatment assignment and outcomes.

In addition to the observed covariates, we use a prognostic score $s_i$ summarizing the baseline risk of patient $i$. The score $s_i$ is obtained either from an externally validated prognostic model or from a predictive model trained on the observed covariates, using cross-fitting on the untreated cohort to avoid overfitting. Throughout the methodology, $s_i$ is treated as an additional baseline feature: it may be used when constructing neighborhoods for the latent factor $U$, and later as a balancing feature when aligning treated and untreated cohorts.

For each patient $i$ and a fixed neighborhood size $k$, we define a directional neighbor set $\mathcal{N}_{i,k}$ restricted to the same treatment arm ($t_j = t_i$) and selected among patients whose observed survival trajectory is opposite in the relevant sense.

\begin{itemize}
    \item If patient $i$ experienced the event ($E_i=1$) at time $T_i$, candidate neighbors are those who survived longer (i.e., had a later event time or were censored after $T_i$). Operationally, we select neighbors among patients satisfying $T_j > T_i$.
    
    \item If patient $i$ was censored ($E_i=0$) at time $T_i$, candidate neighbors are those who experienced the event earlier, i.e., patients satisfying $E_j=1$ and $T_j < T_i$.
\end{itemize}

Among these candidates, we select the $k$ nearest neighbors in the $X$-space according to a distance metric $d(\cdot,\cdot)$ (Euclidean distance after standardizing the covariates). Formally, for event patients ($E_i=1$),

\begin{equation}
\mathcal{N}_{i,k} 
= \operatorname*{arg\,min}_{\substack{\mathcal{S}\subseteq \{j:\, t_j=t_i,\; T_j>T_i\}\\ |\mathcal{S}|=k}}
\sum_{j\in \mathcal{S}} d(X_i,X_j),
\qquad \text{if }E_i=1,
\label{eq:neighbors_event}
\end{equation}

and analogously, for censored patients ($E_i=0$),

\begin{equation}
\mathcal{N}_{i,k} 
= \operatorname*{arg\,min}_{\substack{\mathcal{S}\subseteq \{j:\, t_j=t_i,\; E_j=1,\; T_j<T_i\}\\ |\mathcal{S}|=k}}
\sum_{j\in \mathcal{S}} d(X_i,X_j),
\qquad \text{if }E_i=0.
\label{eq:neighbors_cens}
\end{equation}

This construction ensures that comparisons are both (i) within treatment and (ii) anchored on patients whose observed survival trajectories represent the opposite behavior relative to patient $i$, conditional on observed follow-up information.

\subsubsection{Residual-type latent factor}

We then define the raw latent factor as the difference between patient $i$'s outcome signal and the average outcome signal of its directional neighbors:

\begin{equation}
U_i \;=\; Y_i \;-\; \frac{1}{|\mathcal{N}_{i,k}|}\sum_{j\in \mathcal{N}_{i,k}} Y_j.
\label{eq:u_def}
\end{equation}

This definition admits a natural structural interpretation. Let the outcome signal of interest be written as

\begin{equation}
Y_i = f_t(X_i, V_i) + \epsilon_i,
\label{eq:y_definition}
\end{equation}

\noindent
where $t$ indexes the treatment arm ($t=0$ untreated, $t=1$ treated), $f_t(\cdot)$ captures the joint contribution of observed and unobserved covariates under treatment $t$, and $\epsilon_i$ represents idiosyncratic noise. This functional form is intentionally general and does not impose structural assumptions on the relationship between observed covariates, latent factors, and outcomes.

Substituting \eqref{eq:y_definition} into \eqref{eq:u_def} yields

\begin{align}
U_i &= \left( f_t(X_i, V_i) - \frac{1}{|\mathcal{N}_{i,k}|} \sum_{j \in \mathcal{N}_{i,k}} f_t(X_j, V_j) \right) + \epsilon_i - \bar{\epsilon}_{\mathcal{N}_{i,k}} \nonumber \\
&= \left( f_t(X_i, V_i) - \frac{1}{|\mathcal{N}_{i,k}|} \sum_{j \in \mathcal{N}_{i,k}} f_t(X_i, V_j) \right) \nonumber \\
&\quad + \frac{1}{|\mathcal{N}_{i,k}|} \sum_{j \in \mathcal{N}_{i,k}} \left( f_t(X_i, V_j) - f_t(X_j, V_j) \right) + \epsilon_i - \bar{\epsilon}_{\mathcal{N}_{i,k}}.
\label{eq:u_expand_general}
\end{align}

Because neighbors are selected to be close in the covariate space ($X_j \approx X_i$), the second bracketed term, which represents differences due to observed features, is expected to be small.

Under this local approximation, $U_i$ can therefore be interpreted as a contrast in the unobserved component:

\begin{equation}
U_i \approx f_t(X_i, V_i) - \mathbb{E}[f_t(X_i, V) \mid \mathcal{N}_{i,k}],
\label{eq:u_proxy}
\end{equation}

\noindent
where the expectation is taken over the directional neighbor set. 

Thus, $U_i$ captures how much the realized prognosis of patient $i$ deviates from that of patients with similar observed characteristics receiving the same treatment, which is attributed to unobserved factors. Using a neighborhood rather than a single closest neighbor is important, as averaging across neighbors smooths variation and reduces the influence of extreme outcomes that could otherwise distort the estimation of $U_i$. 

\subsubsection{Choice of \texorpdfstring{$Y_i$}{Y\_i} in survival settings: pseudo-RMST}

In survival settings, the quantity $Y_i$ in Equation \eqref{eq:y_definition} is not directly observed as a scalar due to censoring. A naive option would be to use the event indicator $E_i$, but this discards time information and is highly sensitive to censoring.

Instead, we use the restricted mean survival time (RMST) as the outcome signal because it summarizes the survival curve up to a clinically meaningful horizon $\tau$ and remains robust to censoring when estimated appropriately. We compute individual-level pseudo-observations for RMST using a jackknife construction \cite{andersen2004regression}. Let $\widehat{\mu}$ denote the RMST estimated from the full sample of size $n$, and $\widehat{\mu}_{(-i)}$ the estimate obtained after leaving patient $i$ out. The jackknife pseudo-observation is

\begin{equation}
Y_i \;=\; n\,\widehat{\mu}\;-\;(n-1)\,\widehat{\mu}_{(-i)}.
\label{eq:pseudo_rmst}
\end{equation}

This construction transforms censored survival data into a scalar outcome contribution reflecting how patient $i$ perturbs the estimated survival curve, providing a smoother and more informative signal than the raw follow-up time and event indicator alone.

With the directional neighbor construction above, the sign of $U_i$ has a natural interpretation:

\begin{itemize}
    \item For event patients ($E_i=1$), neighbors are selected among those who survive longer; hence $Y_i$ is typically smaller than the neighbor average, yielding $U_i<0$, which corresponds to higher latent risk.
    
    \item For censored patients, neighbors are selected among those who experience an earlier event; hence $Y_i$ tends to exceed the neighbor average, yielding $U_i>0$, corresponding to a protective latent profile.
\end{itemize}

Thus the sign of $U$ reflects latent risk direction, while its magnitude captures how atypical the outcome signal is relative to nearby patients.

\subsubsection{Percentile normalization across arms}

To make $U$ comparable across treatment arms while preserving its directional interpretation, we apply a signed normalization within each arm, producing $\tilde{U}\in[-1,1]$. Because the raw values of $U$ may be skewed and contain extreme observations, the normalization is performed separately for positive and negative values within each arm using winsorization that limits the influence of outliers while preserving the sign interpretation.

The key comparability assumption is that the directional meaning of $\tilde{U}$ is consistent across treatment arms: more negative values correspond to stronger latent risk, and more positive values correspond to stronger latent protection relative to patients with similar observed characteristics. The signed within-arm normalization places $|\tilde{U}|$ on a comparable, unitless scale across treatment arms, allowing $\tilde{U}$ to be interpreted as a relative deviation from expected prognosis within arm. Under random treatment assignment, latent prognostic factors are expected to be similarly distributed across arms; therefore, after this normalization, the distribution of $\tilde{U}$ should also be approximately balanced across treatment arms in RCT or RCT-like cohorts. 

The resulting normalized quantity $\tilde U$ is subsequently used as an additional covariate in the balancing step between treatment arms, summarizing the latent prognostic component of each patient.

\subsection{Balancing the Treatment Arms}\label{sec:balancing}

After constructing $\tilde U$, the next step is to balance the treated and untreated cohorts. Specifically, we construct a weighted pseudo-cohort in which the distribution of baseline covariates $X$ and the latent feature $\tilde U$ is comparable between treated and untreated patients; this can enable reliable estimation of the treatment effect. To assess the framework's robustness across different balancing procedures, we consider three complementary strategies: prognostic matching followed by sample weighting, entropy balancing, and inverse propensity weighting. 

\subsubsection{Prognostic matching followed by weighting}
\label{sec:progmatch_weighting}

Prognostic matching \cite{bertsimas2024road} reduces observed confounding by (i) stratifying patients into buckets based on a predicted baseline risk score (prognostic score), and (ii) performing one-to-one matching within each bucket using standardized baseline covariates. This design encourages similarity both in baseline covariate profiles and in baseline prognosis.

Although $\tilde U$ could in principle be treated as an additional baseline covariate and included in the stratification, we avoid doing so during the prognostic-matching step because $\tilde U$ is constructed to be strongly associated with the outcome, and would distort the interpretation of the prognostic buckets. Furthermore, prognostic matching primarily enforces similarity in the prognostic score rather than strict similarity in the individual covariates $X$. Treating $\tilde{U}$ as an additional feature when calculating distances for the matching step would therefore not guarantee adequate balance in the effect of the unobserved factors.

Instead, we extend the weighting approach of prognostic matching. After performing prognostic matching, we compute a scalar weight $w$ that is applied to a pre-specified subgroup of the treated cohort in order to align the mean of $\tilde U$ between treated and untreated patients.
More specifically, let $\mathcal{T}=\{i:t_i=1\}$ and $\mathcal{U}=\{i:t_i=0\}$ denote the treated and untreated sets, respectively. We denote by $\mathcal{G}=\{i:E_i=0\}$ the set of patients without an observed event and by $\mathcal{B}=\{i:E_i=1\}$ the set of patients with an observed event. In our implementation, we apply this scalar weight to the subgroup $\mathcal{T}\cap\mathcal{G}$ (treated patients without an observed event), which yields a simple and interpretable reweighted treated cohort. We note that one could alternatively assign weights to other subgroups. We use a single scalar weight to avoid instability and overfitting that may arise from highly flexible, individual-specific weights.

Concretely, we calculate $w$ such that the mean of $\tilde U$ among the untreated patients equals the mean of $\tilde U$ in the reweighted treated cohort:
\begin{equation}
\label{eq:w_scalar_balanceU}
\frac{1}{|\mathcal{U}|}\sum_{i\in\mathcal{U}}\tilde U_i
=
\frac{
\sum_{i\in\mathcal{T}\cap\mathcal{B}}\tilde U_i
+
w\sum_{i\in\mathcal{T}\cap\mathcal{G}}\tilde U_i
}{
|\mathcal{T}\cap\mathcal{B}|
+
w|\mathcal{T}\cap\mathcal{G}|
}.
\end{equation}
Solving \eqref{eq:w_scalar_balanceU} for $w$ yields a reweighted treated cohort whose mean $\tilde U$ matches that of the untreated cohort. To avoid extreme weights and to ensure positivity, we constrain $w\in[0.5,20]$; in practice, extreme values are rare.


\subsubsection{Entropy balancing}
\label{sec:entropy_balancing}

As a second balancing approach, we use entropy balancing to reweight control patients so that selected moments of certain covariates match those of the treated cohort. Specifically, we compute nonnegative weights for control patients that approximately match the first moments of the treated group for the baseline covariates, the prognostic score, and the latent factor $\tilde U$. In addition, we match the second moments of the latter two quantities to better account for differences in their distributions across treatment arms, as the prognostic score summarizes observed features and $\tilde U$ summarizes latent prognostic factors.

Let $Z_i$ denote the vector of features to be balanced. In our implementation, 
\[
Z_i = \big[X_i,\; s_i,\; \tilde U_i,\; s_i^2,\; \tilde U_i^2 \big],
\]
where $X_i$ are standardized baseline covariates and $s_i$ is the prognostic score.

Entropy balancing assigns weights $w_i$ to controls ($t_i=0$) by solving the convex optimization problem
\begin{align}
\label{eq:entropy_primal}
\min_{\{w_i\}_{i\in\mathcal{U}}}\quad & \sum_{i\in\mathcal{U}} w_i \log w_i \\
\text{s.t.}\quad 
& \frac{1}{|\mathcal{T}|}\sum_{i\in\mathcal{T}} Z_i
=
\frac{1}{|\mathcal{T}|}\sum_{i\in\mathcal{U}} w_i Z_i, \nonumber\\
& \sum_{i\in\mathcal{U}} w_i = |\mathcal{T}|,\quad w_i\ge 0,\ \forall i\in\mathcal{U}. \nonumber
\end{align}
We set weights for treated patients to one ($w_i=1$ for $i\in\mathcal{T}$) and only reweight the controls. 


\subsubsection{Inverse propensity weighting (IPTW)}
\label{sec:iptw}

The third balancing method is inverse propensity weighting, aimed at estimating weights for the untreated cohort so that it resembles the treated cohort. We fit a logistic regression propensity model on the full cohort to estimate $p_i = \mathbb{P}(t_i=1 \mid X_i, s_i, \tilde U_i)$. We use logistic regression (rather than more flexible models) to reduce overfitting and improve stability.

We then apply the weights by assigning weight $1$ to treated patients and reweighting controls according to
\begin{equation}
\label{eq:iptw_att}
w_i =
\begin{cases}
1, & t_i = 1,\\
\dfrac{p_i}{1-p_i}, & t_i = 0.
\end{cases}
\end{equation}
This weighting scheme upweights control patients who look more likely to have been treated, making the weighted control cohort comparable to the treated cohort, similarly to the entropy balancing method.

\subsection{Multivariable analysis}
\label{sec:cox_adjustment}

After balancing, we estimate the treatment effect using a multivariable Cox proportional hazards model fit on the weighted samples, with the weights determined by the chosen balancing strategy. The Cox model includes the treatment indicator and all baseline observed covariates (and prognostic score when applicable). We do not include $\tilde U$ as a covariate in this final regression. The influence of $\tilde U$ is already incorporated through the balancing weights; including $\tilde U$ directly in the Cox model would introduce instability due to the strong association between $\tilde U$ and the outcome. The coefficient of the treatment indicator from the weighted Cox model is reported as our adjusted estimate of treatment effect. 

Taken together, these three steps produce balanced cohorts with respect to both observed covariates and the latent prognostic feature $\tilde U$, enabling more reliable estimation of treatment effects in observational cohorts.

\subsection{Statistical analysis of validation experiments}\label{sec:validation_stats}

To evaluate our framework and its robustness across different settings, we experimented with multiple hyperparameter configurations; all tuning ranges were defined \emph{a priori}, and for each balancing method the baseline and $\tilde{U}$-augmented versions were run over identical hyperparameter grids. We then summarized performance across the full set of pre-specified parameter combinations (rather than selecting the best-performing setting), which reduces the risk that apparent gains of the augmented approach are attributable to post hoc hyperparameter selection.

For descriptive visualization, we report average HRs and standard errors across parameter combinations (Tables~\ref{tab:obs_val_1}--\ref{tab:crlm_comp}). Formal inference was based on paired comparisons between the corresponding baseline and $\tilde{U}$-augmented analyses within matched settings (same dataset, same balancing method, and same hyperparameter configuration).

For Validation 1 (observational cohorts with benchmark RCT HRs), the performance metric was absolute distance to the benchmark on the log-HR scale. For each matched run $c$, we computed
\[
d_c^{\mathrm{base}}=\left|\log\!\left(HR_c^{\mathrm{base}}\right)-\log\!\left(HR^{\mathrm{RCT}}\right)\right|,\] \\
\[
d_c^{\mathrm{aug}}=\left|\log\!\left(HR_c^{\mathrm{aug}}\right)-\log\!\left(HR^{\mathrm{RCT}}\right)\right|,
\]
and defined the paired improvement
\[
\Delta_c = d_c^{\mathrm{base}} - d_c^{\mathrm{aug}},
\]
so that $\Delta_c>0$ indicates that adding $\tilde{U}$ moved the estimate closer to the benchmark RCT HR. To avoid treating hyperparameter combinations as independent observations, we first summarized $\Delta_c$ within each dataset-by-balancing-method cell (mean across parameter combinations), then tested whether these cell-level summaries were systematically positive using an exact sign test (primary analysis). Ties (cell summaries exactly equal to zero) were excluded from the sign-test denominator. A Wilcoxon signed-rank test on the same cell-level summaries was used as a sensitivity analysis. As an effect-size summary, we report the median and mean cell-level improvement in benchmark distance on the log-HR scale (Table \ref{tab:obs_val_1}).

For Validation 2 (RCT cohorts), the objective was to verify that adding $\tilde{U}$ did not materially distort treatment-effect estimates in settings where confounding should be absent by design. We therefore used equivalence testing. The augmentation-induced shift was defined as
\[
\Delta_{\mathrm{shift}}=\log\!\left(HR^{\mathrm{aug}}\right)-\log\!\left(HR^{\mathrm{RCT}}\right),
\]
and evaluated using two one-sided tests (TOST) against a pre-specified equivalence margin $\delta$, testing whether $-\delta < \Delta_{\mathrm{shift}} < \delta$. The equivalence margin was specified \emph{a priori} as a clinically negligible relative change in HR (e.g. $\delta=\log(1.10)$). Equivalence was concluded only if both one-sided tests were significant. We additionally report the observed log-HR shifts as effect-size summaries (Table \ref{tab:rct_summary}).

For Validation 3 (cross-center consistency in CRLM), we consider two complementary analyses corresponding to distinct estimands: (i) the consistency of treatment-effect estimates across centers, and (ii) differences in survival outcomes driven by variation in patient case mix.

\paragraph{(i) Treatment-effect consistency across centers.}
Between-center heterogeneity in treatment-effect estimates was quantified using the mean absolute pairwise difference of center-specific treatment effects on the log-HR scale. For each matched setting $c$ (same outcome, same balancing method, and same hyperparameter configuration), we computed $D_c^{\mathrm{base}}$ and $D_c^{\mathrm{aug}}$, where

\[
D_c = \frac{1}{\binom{C}{2}} \sum_{i<j}
\left| \log(HR_{i,c}) - \log(HR_{j,c}) \right|,
\]

and defined

\[
\Delta D_c = D_c^{\mathrm{base}} - D_c^{\mathrm{aug}},
\]

\noindent
so that positive values indicate reduced between-center variability after incorporating $\tilde{U}$. This analysis targets within-center confounding: if latent confounders are shared across centers and captured by $\tilde{U}$, adjustment should yield more comparable HR estimates.

As in Validation 1, we summarized $\Delta D_c$ within each outcome-by-balancing-method cell (mean across parameter combinations) and tested whether the cell-level summaries were systematically positive using an exact sign test (primary analysis), with a Wilcoxon signed-rank test as a sensitivity analysis. Ties were excluded from the sign-test denominator. As an effect-size summary, we report the median and mean cell-level reduction in pairwise dispersion on the log-HR scale (Table~\ref{tab:crlm_comp}).

\paragraph{(ii) Cross-center survival comparisons.}
For the CRLM centers, we performed an additional validation experiment to assess the role of $\tilde{U}$ beyond $X$ when comparing adjusted 5-year survival across centers. In contrast to the HR analysis, this setting does not target confounding in treated-versus-untreated comparisons, but rather differences in patient selection (case mix) across centers. Accordingly, instead of balancing treated and untreated arms within each center, we use the same balancing methods to construct pairwise cross-center comparisons by reweighting patients from a source center to match the covariate and latent-factor distribution of a target center. This allows survival differences to be interpreted conditional on comparable patient populations across centers.

For each pair of centers $(A,B)$, suppose without loss of generality that center $A$ has lower crude 5-year survival than center $B$. We define
\[
D_{AB}^{\mathrm{raw}} = S_B^{\mathrm{raw}}(5)-S_A^{\mathrm{raw}}(5),\]
\[
D_{AB}^{\mathrm{base}} = S_B^{\mathrm{base}}(5)-S_A^{\mathrm{base}}(5),
\]
\[
D_{AB}^{\mathrm{aug}} = S_B^{\mathrm{aug}}(5)-S_A^{\mathrm{aug}}(5),
\]
where $S_A^{\mathrm{raw}}(5)$, $S_A^{\mathrm{base}}(5)$, and $S_A^{\mathrm{aug}}(5)$ denote the crude, $X$-adjusted, and $(X,\tilde U)$-adjusted 5-year survival estimates for center $A$, respectively.

We focused on the subset of center pairs for which the ordering was preserved after adjustment on $X$ (i.e., $S_A^X(5)<S_B^X(5)$) and for which adjustment using only observed covariates widened the survival gap, namely
\[
D_{AB}^{X} > D_{AB}^{\mathrm{raw}}.
\]
This pattern is informative: if balancing on observed characteristics makes the patient populations more comparable, yet the survival disadvantage of center $A$ becomes larger, it suggests the presence of important unobserved prognostic differences between centers.

Since $\tilde U$ is designed to capture such latent prognostic variation, we would expect that adjusting for $\tilde U$ in addition to $X$ reduces this enlarged gap, that is,
\[
D_{AB}^{\mathrm{aug}} < D_{AB}^{\mathrm{base}}.
\]

For this validation experiment, we compute the percentage of selected center pairs for which adjustment using $(X,\tilde U)$ reduces the $X$-adjusted survival gap. A higher percentage indicates that $\tilde U$ captures previously unobserved prognostic differences between centers. We then perform an exact binomial test against the null hypothesis of success probability $0.5$. 

We note that this analysis targets a diagnostic subset of center pairs for which adjustment on $X$ reveals potential unobserved prognostic differences; for other center pairs, the addition of $\tilde{U}$ may either increase or decrease the survival gap depending on the relative contributions of observed and latent prognostic factors.

All analyses were performed in Python (version 3.8) using standard scientific computing libraries, including NumPy, pandas, scikit-learn, lifelines, and SciPy. Statistical tests were conducted using the scipy.stats module.

\section{Results}

In this Section, we present the results of our experiments evaluating how incorporating the latent factor $\tilde{U}$ during the balancing step influences downstream estimates, including both treatment effects (hazard ratios) and cross-center survival comparisons. We conducted three groups of experiments designed to assess performance under complementary settings regarding the types of datasets employed.

To estimate $\tilde{U}$, several hyperparameters must be specified. We explored multiple design choices and report average results across all configurations. These include: (i) the number of nearest neighbors $k$ used to compute $\tilde{U}_i$ for anchor patient $i$, (ii) whether the prognostic score $s_i$ is included in the distance metric for neighbor selection, and (iii) which prognostic score is utilized when applicable. 

Each balancing method also involves its own design parameters. For prognostic matching, we vary the number of risk strata used during stratification. For entropy balancing, we examine whether matching only first moments or both first and second moments of $\tilde{U}_i$ and the prognostic score $s_i$ affects performance. For IPTW, we vary clipping thresholds to mitigate extreme weights. An extended sensitivity analysis for all hyperparameters is provided in Appendix Section \ref{app_sect:sens_analysis}.

\subsection{Real-World Observational Data}

In the first set of validations, we applied our framework to observational oncology cohorts. The objective was to determine whether the derived cohorts produce treatment-effect estimates that align with ground-truth HRs reported in RCTs evaluating the same therapies.

For each dataset, we estimated the HR of the treatment within the constructed cohort and compared it with literature-established HRs from prior RCTs. The HR was computed as the exponentiated coefficient of the treatment indicator from a Cox proportional hazards model.

We evaluated our approach on three observational oncology datasets with tumor recurrence as the event of interest. Table~\ref{tab:cohort_summary_obs} summarizes cohort size and treatment/outcome distributions.

For each dataset, we used a clinically established feature set $X_i$, a prognostic score $s_i$, and the latent feature $U_i$ to define the profile of patient $i$. We examined multiple prognostic scores: an externally validated score when available, and an internally trained model fitted on untreated patients only, predicting event probability within a fixed time horizon using cross-fitting.

\begin{table*}[ht]
\centering
\caption{Cohort characteristics of the observational datasets used in the study.}
\label{tab:cohort_summary_obs}
\begin{tabular}{l c c c c c}
\toprule
Dataset & N & Surgery + adjuvant therapy & Surgery alone & Events (Surgery + adjuvant therapy) & Events (Surgery alone) \\
\midrule
GIST & 480 
& 105 (21.9\%) & 375 (78.1\%) 
& 34 (32.4\%) & 52 (13.9\%) \\

RPS & 201 
& 29 (14.4\%) & 172 (85.6\%) 
& 3 (10.3\%) & 49 (28.5\%) \\

CRLM & 778 
& 140 (18.0\%) & 638 (82.0\%) 
& 83 (59.3\%) & 445 (69.8\%) \\

\bottomrule
\end{tabular}
\end{table*}

The Memorial Sloan Kettering Cancer Center (MSKCC) gastrointestinal stromal tumor (GIST) dataset comprises patients with resected GIST treated with or without adjuvant imatinib; details of this cohort have been reported previously \cite{bertsimas2024road}. According to the benchmark RCT, adjuvant imatinib was associated with an HR of 0.35 for recurrence-free survival \cite{dematteo2009adjuvant}. Our feature set included the most commonly used prognostic factors in GIST risk models, namely tumor size, tumor site, and mitotic index \cite{bertsimas2023interpretable}. The external prognostic model used in our analysis \cite{gold2009development} incorporates these same variables.

The MSKCC retroperitoneal sarcoma (RPS) dataset includes patients with retroperitoneal or pelvic sarcoma treated with surgery alone or surgery plus perioperative radiotherapy; details of this cohort have been reported previously \cite{kelly2015comparison}. According to the per-protocol analysis of the benchmark RCT, pre-operative radiotherapy was associated with an HR of 0.75 for abdominal recurrence-free survival \cite{bonvalot2020preoperative}. Our feature set included the most commonly used prognostic factors in RPS local recurrence risk models, namely age, tumor size, grade, histology, and margin status. The external prognostic model used in our analysis similarly incorporates these same variables \cite{bertsimas2023interpretable, kelly2015comparison}.

The CRLM multi-institutional dataset comprised patients with resected colorectal liver metastases treated with or without adjuvant chemotherapy after surgery; these patients were selected to match the inclusion criteria of the most recent RCT evaluating adjuvant chemotherapy in CRLM. Details of this cohort have been reported previously \cite{bertsimas2025road}. According to that benchmark RCT, adjuvant FOLFOX chemotherapy was associated with an HR of 0.67 for disease-free survival \cite{kanemitsu2021hepatectomy}. The external prognostic model used in our analysis incorporates age, sex, primary tumor T stage, lymph node status (N stage), KRAS mutational status, tumor sidedness, CEA level, disease-free interval, tumor size, bilobar involvement, and number of metastases \cite{kawaguchi2021contour}.

\begin{figure*}[ht]
\centering
\includegraphics[width=0.9\linewidth]{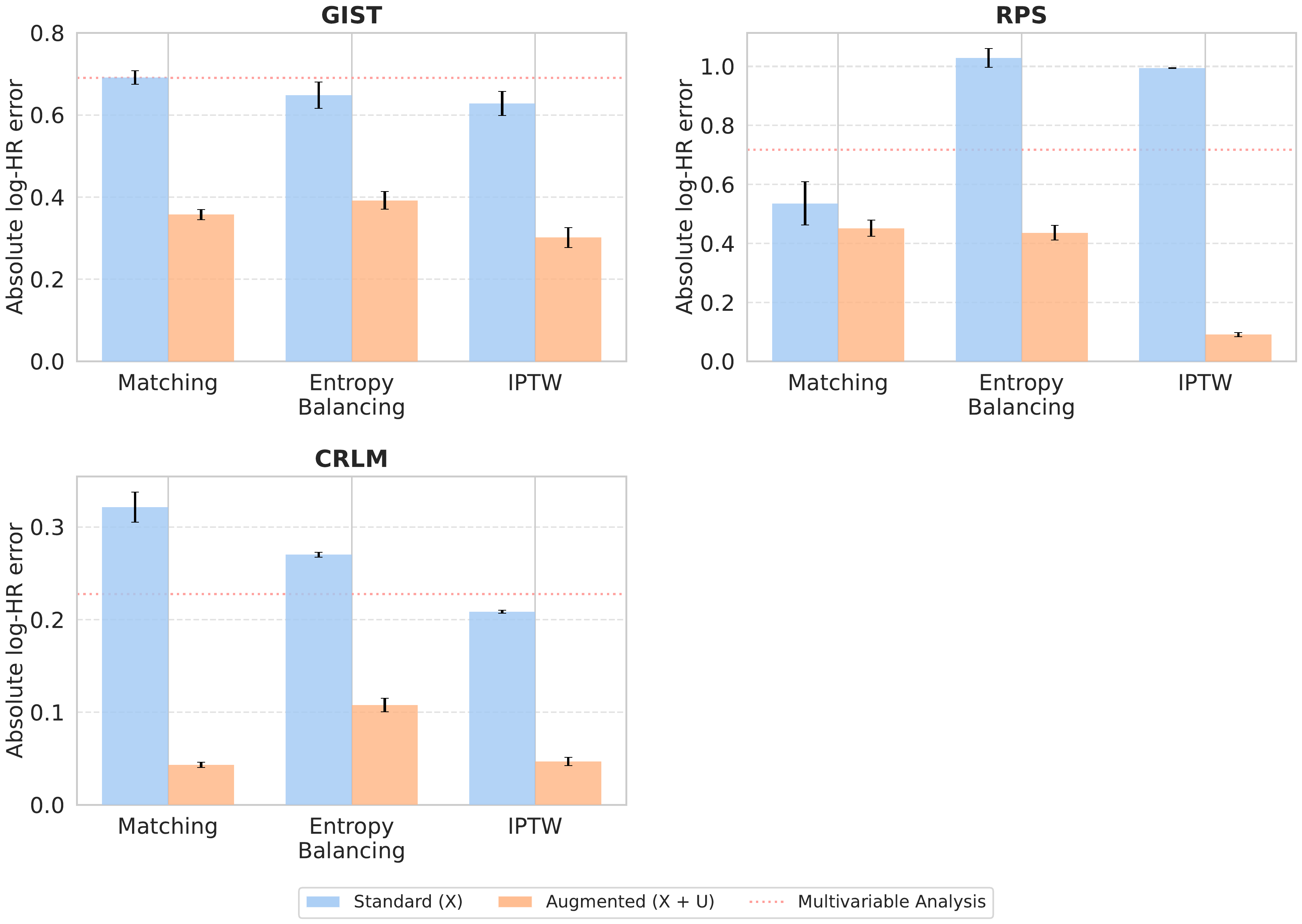}
\caption{Comparison of treatment-effect estimation accuracy across observational datasets. Bars represent the mean absolute log-HR error relative to the benchmark RCT hazard ratio, averaged across hyperparameter configurations. Results are shown for matching, entropy balancing, and inverse propensity weighting using either baseline covariates $X$ or the augmented representation $(X,\tilde{U})$. The dashed line denotes multivariable Cox adjustment using baseline covariates only. Error bars represent the standard error across configurations.}
\label{fig:obs_loghr_plot}
\end{figure*}

To quantify agreement with the benchmark RCT estimates, we measure the absolute
log-HR error between the estimated hazard ratio and the RCT reference value.
For dataset $d$, balancing method $m$, variant $v$ ($X$-adjustment or $(X, \tilde{U})$-adjustment), and hyperparameter
configuration $h$, the error is defined as

\begin{equation}
\text{Error}_{d,m,v,h} =
\left|
\log(\widehat{\mathrm{HR}}_{d,m,v,h}) -
\log(\mathrm{HR}^{\mathrm{RCT}}_{d})
\right|.
\end{equation}

The plotted value corresponds to the mean error across all hyperparameter
configurations:
\begin{equation}
\overline{\text{Error}}_{d,m,v} =
\frac{1}{H}
\sum_{h=1}^{H}
\text{Error}_{d,m,v,h}.
\end{equation}

Figure \ref{fig:obs_loghr_plot} presents these average log-HR errors across all configurations for simple covariate adjustment as a baseline, compared with prognostic matching, entropy balancing, and inverse propensity weighting. Across all three observational datasets and across all balancing methods, incorporating the latent factor $\tilde{U}$ consistently reduces the estimation error.

\begin{figure*}[ht]
\centering
\includegraphics[width=0.9\linewidth]{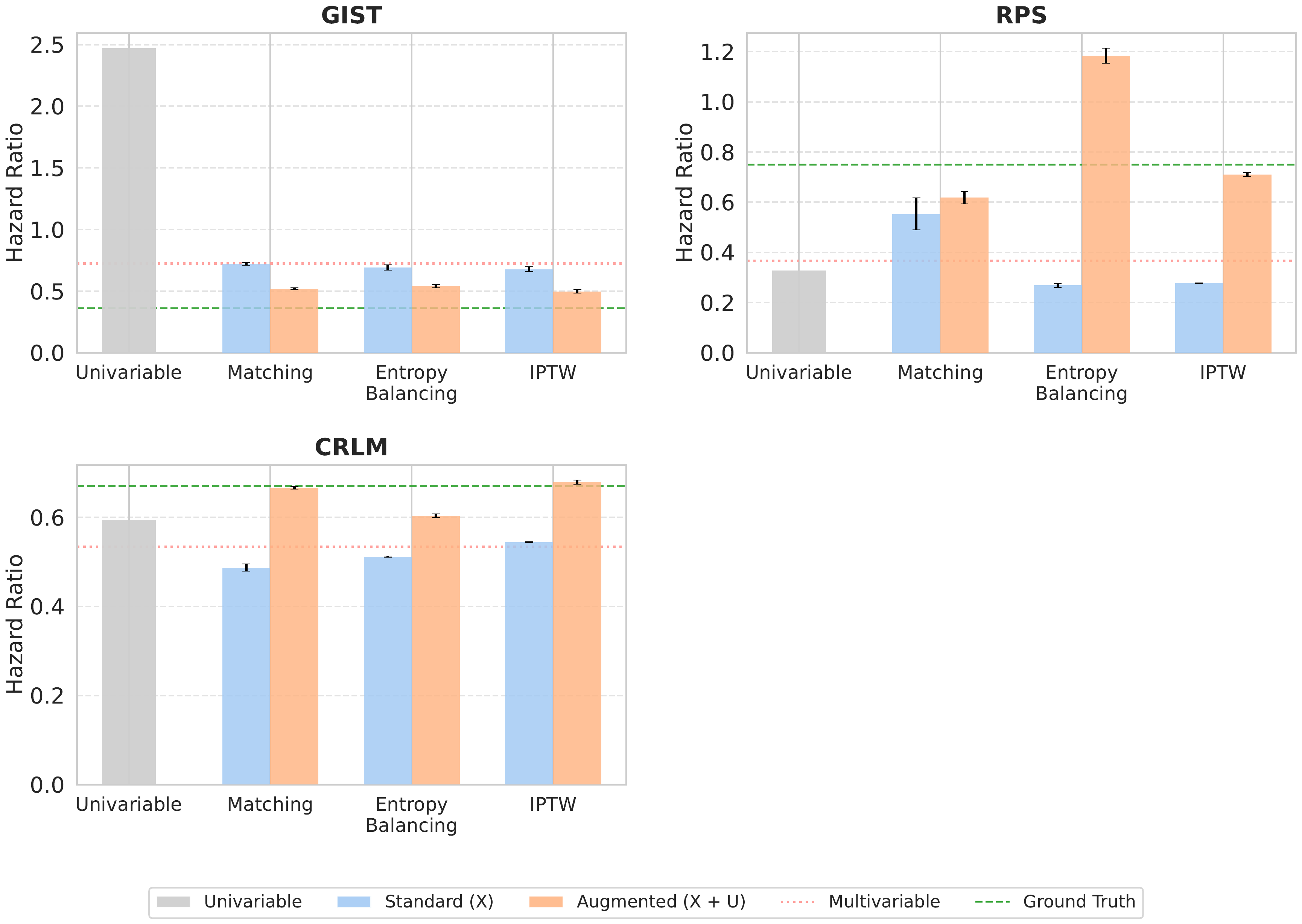}
\caption{Comparison of treatment-effect estimates across observational datasets. Bars represent the mean hazard ratio (HR), averaged across hyperparameter configurations. Results are shown for matching, entropy balancing, and inverse propensity weighting using either baseline covariates $X$ or the augmented representation $(X,\tilde{U})$. The dotted line denotes multivariable Cox adjustment using baseline covariates only, and the dashed green line indicates the benchmark RCT hazard ratio. Error bars represent the standard error across configurations.}
\label{fig:obs_hr_plot}
\end{figure*}

The raw HR values are presented in Figure \ref{fig:obs_hr_plot}. The only dataset that exhibits instability is RPS. This is expected, as the RPS cohort is relatively small and contains very few events in the treated group, making reliable HR estimation more difficult. Despite this limitation, all methods incorporating $\tilde{U}$ improve upon adjustment using $X$ alone. Among the evaluated approaches, IPTW generally produces the most stable estimates. Entropy balancing occasionally overshoots the benchmark HR, likely due to extreme weights produced by the balancing constraints.

To quantify whether incorporating $\tilde{U}$ systematically improves agreement with the benchmark RCT hazard ratios, we evaluated the directional improvement in absolute log-HR error between models using $X$ alone and those using $(X,\tilde{U})$. Statistical testing procedures are described in Section~\ref{sec:validation_stats}. Across all datasets and balancing methods, incorporating $\tilde{U}$ reduced the error relative to the RCT benchmark, resulting in consistent directional improvements (9/9 comparisons). As summarized in Table~\ref{tab:obs_val_1}, this improvement was statistically significant under both the sign test and the Wilcoxon signed-rank test. The differences for all pairs of datasets and balancing methods can be found in the Appendix Table \ref{tab:method_comparison}.

\begin{table}[htbp]
\centering
\caption{Validation approach 1: directional improvement in closeness to the benchmark RCT hazard ratio after adding $\tilde{U}$.}
\label{tab:obs_val_1}
\small
\begin{tabular}{lr}
\toprule
Statistic & Value \\
\midrule
Directional improvements ($\Delta>0$) & 9 / 9 \\
Sign test $p$-value (one-sided) & 0.00195 \\
Wilcoxon signed-rank $p$-value (one-sided) & 0.00195 \\
Median cell-level $\Delta$ & 0.2782 \\
Mean cell-level $\Delta$ & 0.3443 \\
\bottomrule
\end{tabular}
\end{table}

\subsection{Randomized Controlled Trials}

In the second set of validations, we applied our methodology to RCT datasets. 
Because treatment assignment in RCTs is randomized, both observed and unobserved confounders are expected to be balanced between study arms. 
Therefore, incorporating $\tilde{U}$ in the balancing methods should not materially alter the estimated HR. 
In contrast to the observational validation, where estimates are compared against benchmark RCT results, the purpose of this validation is to verify that incorporating the latent factor $\tilde{U}$ does not materially change treatment-effect estimates when treatment assignment is already randomized.

We evaluated two RCT datasets: STRASS and BRT.

The STRASS RCT\cite{bonvalot2020preoperative} compared preoperative radiotherapy plus surgery versus surgery alone in patients with RPS, with abdominal recurrence-free survival as the primary trial endpoint. The reported HR in the intention-to-treat analysis was 1.00, indicating no treatment benefit. The prognostic features used in our analysis were age, tumor size, tumor grade and histological subtype.


The BRT RCT included patients with soft tissue sarcomas of the extremity or superficial trunk treated with either adjuvant brachytherapy (BRT) or no further therapy after complete resection, with local recurrence as the primary trial endpoint \cite{pisters1996long}. The prognostic features used in our analysis were tumor histology, margin status, tumor grade, tumor size, and tumor depth.


Summary statistics are provided in Table~\ref{tab:summary_rct}.

\begin{table*}[ht]
\centering
\caption{Cohort characteristics of the RCT datasets.}
\label{tab:summary_rct}

\setlength{\tabcolsep}{4pt}

\begin{tabular}{l c c c c c}
\toprule
& \multicolumn{2}{c}{Treatment} & \multicolumn{2}{c}{Events} \\
\cmidrule(lr){2-3} \cmidrule(lr){4-5}
Dataset & N & Surgery + periop. therapy & Surgery alone & Surgery + periop. therapy & Surgery alone \\
\midrule
BRT & 162 & 75 (46.3\%) & 87 (53.7\%) & 12 (16.0\%) & 26 (29.9\%) \\
STRASS & 265 & 132 (49.8\%) & 133 (50.2\%) & 59 (44.7\%) & 61 (45.9\%) \\
\bottomrule
\end{tabular}
\end{table*}

\begin{figure*}[htbp!]
\centering
\includegraphics[width=0.9\linewidth]{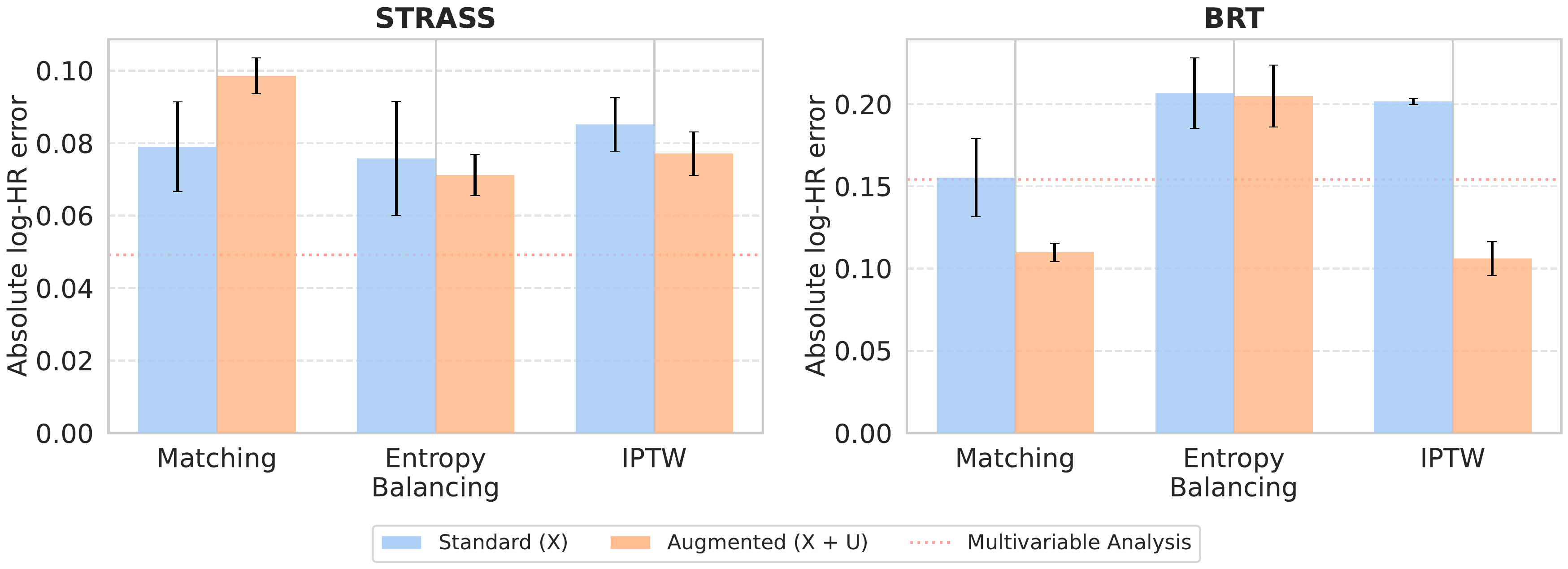}
\caption{Comparison of treatment-effect estimation accuracy across RCT datasets. Bars represent the mean absolute log-HR error relative to the trial-reported hazard ratio, averaged across hyperparameter configurations. Results are shown for matching, entropy balancing, and inverse propensity weighting using either baseline covariates $X$ or the augmented representation $(X,\tilde{U})$. The dotted line denotes multivariable Cox adjustment using baseline covariates only. Error bars denote the standard error across configurations.}
\label{fig:rct_log_hr_plot}
\end{figure*}


\begin{figure*}[htbp!]
\centering
\includegraphics[width=0.9\linewidth]{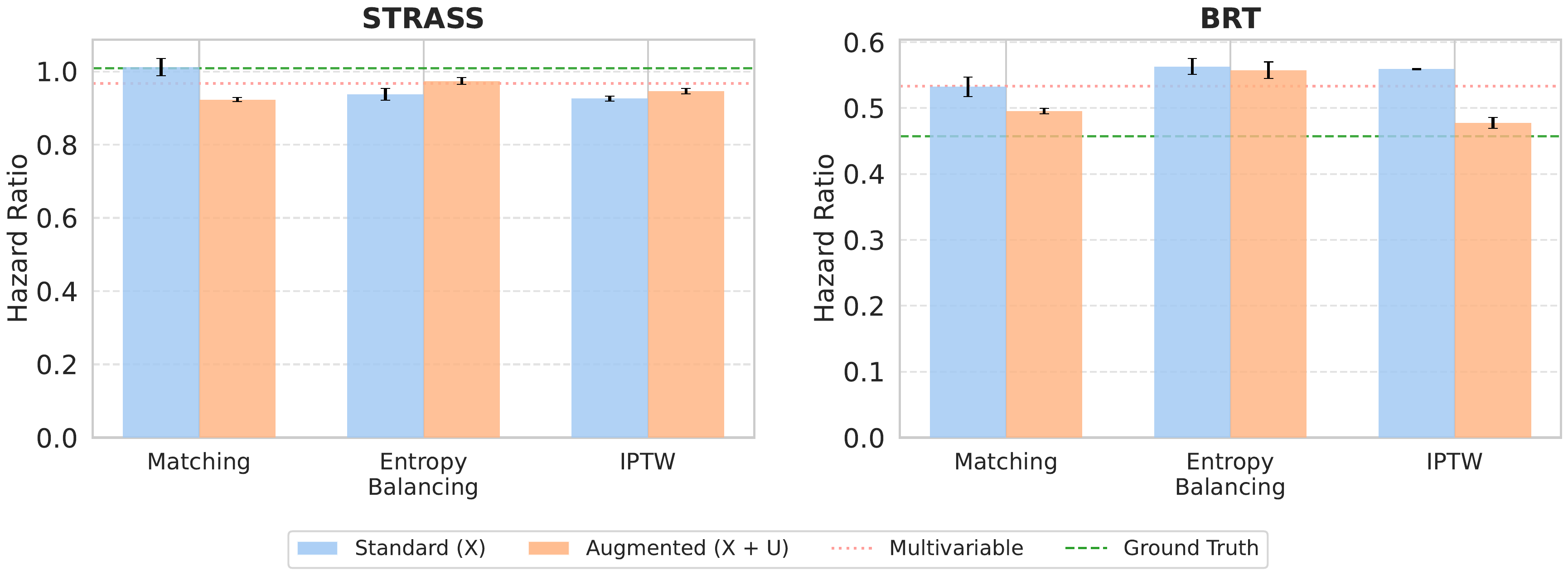}
\caption{Comparison of treatment-effect estimates across RCT datasets. Bars represent the mean hazard ratio (HR), averaged across hyperparameter configurations, for matching, entropy balancing, and inverse propensity weighting using either baseline covariates $X$ or the augmented representation $(X,\tilde{U})$. Error bars denote the standard error across configurations. The dotted line indicates the multivariable estimate, and the dashed green line indicates the trial-reported ground-truth HR.}
\label{fig:rct_hr_plot}
\end{figure*}

\begin{table}[ht]
\centering
\caption{Validation approach 2: equivalence testing for RCT datasets. A ``win'' indicates that the augmentation with $\tilde{U}$ passes the equivalence test, meaning the 95\% confidence interval of the shift $\Delta_{\text{shift}} = \log(HR^{\mathrm{aug}})-\log(HR^{\mathrm{RCT}})$ lies entirely within the pre-specified equivalence margin ($\pm \log(1.10)$).}
\label{tab:rct_summary}
\begin{tabular}{lc}
\toprule
Metric & Value \\
\midrule
Equivalence wins & 4 / 6 \\
Mean $|\Delta_{\text{shift}}|$ & 0.081 \\
\bottomrule
\end{tabular}
\end{table}

As expected, incorporating $\tilde{U}$ in the balancing step does not materially change the HR estimates relative to those obtained using $X$ alone, as we observe in Figures \ref{fig:rct_log_hr_plot} and \ref{fig:rct_hr_plot}. In smaller samples, such as BRT, simple covariate adjustment may yield deviations from the reported HR due to sampling variability; however, the inclusion of $\tilde{U}$ reduces the error relative to the benchmark HR in two of the three balancing methods. This supports the interpretation that $\tilde{U}$ corrects latent imbalance only when such imbalance exists.

The two-sided statistical test described in Section~\ref{sec:validation_stats} was applied using an equivalence margin of $\log(1.10)$, corresponding to a maximum allowable change of approximately 10\% in the hazard ratio after incorporating $\tilde{U}$. Specifically, we evaluated the shift
\[
\Delta_{\text{shift}} = \log(HR^{\mathrm{aug})} - \log(HR^{\mathrm{RCT}}),
\]
and assessed whether the 95\% confidence interval of $\Delta_{\text{shift}}$ lies entirely within the interval $[-\log(1.10), \log(1.10)]$. Using this criterion, equivalence was achieved in four of the six dataset–method combinations. In STRASS, one configuration narrowly exceeded the predefined equivalence margin, while in BRT the most deviating estimate remained close to the corresponding $X$-only adjustment. All corresponding values are reported in the Appendix Table \ref{tab:equivalence_loghr}.

\begin{figure*}[ht]
\centering
\includegraphics[width=0.9\linewidth]{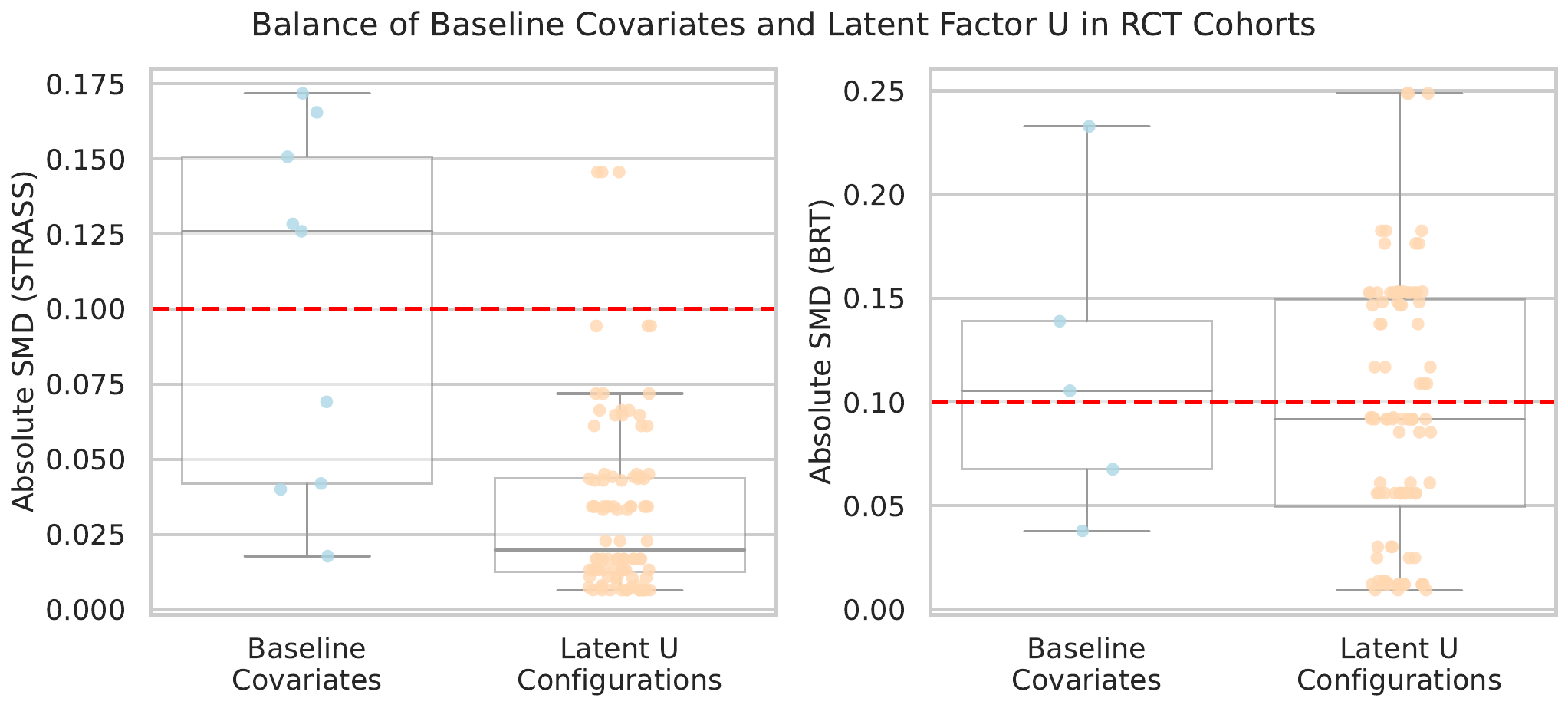}
\caption{Standardized mean differences (SMDs) between treatment arms for observed covariates and estimated latent factors $\tilde{U}$ in the RCT datasets. The SMD values for $\tilde{U}$ are generally below or close to the commonly used balance threshold of 0.1 and are comparable to those of the observed covariates.}
\label{fig:rct_smds}
\end{figure*}

An additional validation examines whether the estimated $\tilde{U}$ differs between treatment arms in the RCT datasets. Because treatment assignment is randomized, both observed covariates and latent prognostic factors are expected to be balanced between arms. To assess this, we computed standardized mean differences (SMDs) for both the observed covariates and the estimated $\tilde{U}$ values across treatment groups (Figure~\ref{fig:rct_smds}). Across both datasets, the majority of SMDs for $\tilde{U}$ are below or close to the commonly used balance threshold of 0.1 and are comparable to, or even smaller than, those observed for the baseline covariates. 

Together, these results confirm that the latent factor $\tilde{U}$ behaves as expected in randomized settings, remaining balanced across treatment arms and not introducing systematic shifts in treatment-effect estimates.

\subsection{Cross-Center Consistency Analysis}

The third set of validations examined six observational CRLM cohorts from six academic centers; details of these cohorts have been reported previously by our group \cite{olthof2022kras}. These cohorts span the USA, Europe, and Asia and are therefore expected to differ in patient selection for surgery, eligibility for adjuvant chemotherapy, and treatment protocols (e.g., chemotherapy regimen). 

Summary statistics are shown in Table~\ref{tab:crlm_centers_summary}. The heterogeneity in both treatment allocation and outcomes across centers is evident. 

The first component of the cross-center validation examines whether balancing on $(X,\tilde{U})$ between adjuvant treatment arms within each center reduces between-center differences in treatment-effect estimates. If a meaningful component of the observed heterogeneity is driven by differences in the distribution of observed and unobserved confounders across centers, then this balancing should reduce inter-center variability in HR estimates. However, even if unmeasured prognostic factors are successfully adjusted for, we would not expect identical HRs for adjuvant chemotherapy across all cohorts, because both treatment exposure (e.g., chemotherapy regimen and treatment duration) and the characteristics of the treated populations may differ between centers.

\begin{table*}[ht]
\centering
\caption{Cohort characteristics of CRLM multi-center datasets.}
\label{tab:crlm_centers_summary}
\resizebox{0.99\textwidth}{!}{
\begin{tabular}{l l c c c c c}
\toprule
Dataset & Event & N & Surgery + chemotherapy & Surgery alone & Events (Surgery + chemotherapy) & Events (Surgery alone) \\
\midrule
Graz and Vienna Universities & Death & 128 & 88 (68.8\%) & 40 (31.2\%) & 47 (53.4\%) & 12 (30.0\%) \\
Graz and Vienna Universities & Recurrence & 128 & 88 (68.8\%) & 40 (31.2\%) & 68 (77.3\%) & 27 (67.5\%) \\
Johns Hopkins & Death & 461 & 296 (64.2\%) & 165 (35.8\%) & 160 (54.0\%) & 116 (70.3\%) \\
Johns Hopkins & Recurrence & 461 & 296 (64.2\%) & 165 (35.8\%) & 183 (61.8\%) & 78 (47.3\%) \\
Kumamoto University & Death & 90 & 46 (51.1\%) & 44 (48.9\%) & 23 (50.0\%) & 26 (59.1\%) \\
Kumamoto University & Recurrence & 90 & 46 (51.1\%) & 44 (48.9\%) & 39 (84.8\%) & 38 (86.4\%) \\
University of Bergen & Death & 220 & 86 (39.1\%) & 134 (60.9\%) & 30 (34.9\%) & 108 (80.6\%) \\
University of Bergen  & Recurrence & 220 & 86 (39.1\%) & 134 (60.9\%) & 48 (55.8\%) & 112 (83.6\%) \\
Yokohama University& Death & 176 & 64 (36.4\%) & 112 (63.6\%) & 27 (42.2\%) & 40 (35.7\%) \\
Yokohama University & Recurrence & 176 & 64 (36.4\%) & 112 (63.6\%) & 51 (79.7\%) & 81 (72.3\%) \\
Cleveland Clinic & Death & 338 & 218 (64.5\%) & 120 (35.5\%) & 81 (37.2\%) & 52 (43.3\%) \\
Cleveland Clinic & Recurrence & 338 & 218 (64.5\%) & 120 (35.5\%) & 156 (71.6\%) & 70 (58.3\%) \\
\bottomrule
\end{tabular}
}
\end{table*}

\begin{figure*}[htbp!]
\centering
\includegraphics[width=0.9\linewidth]{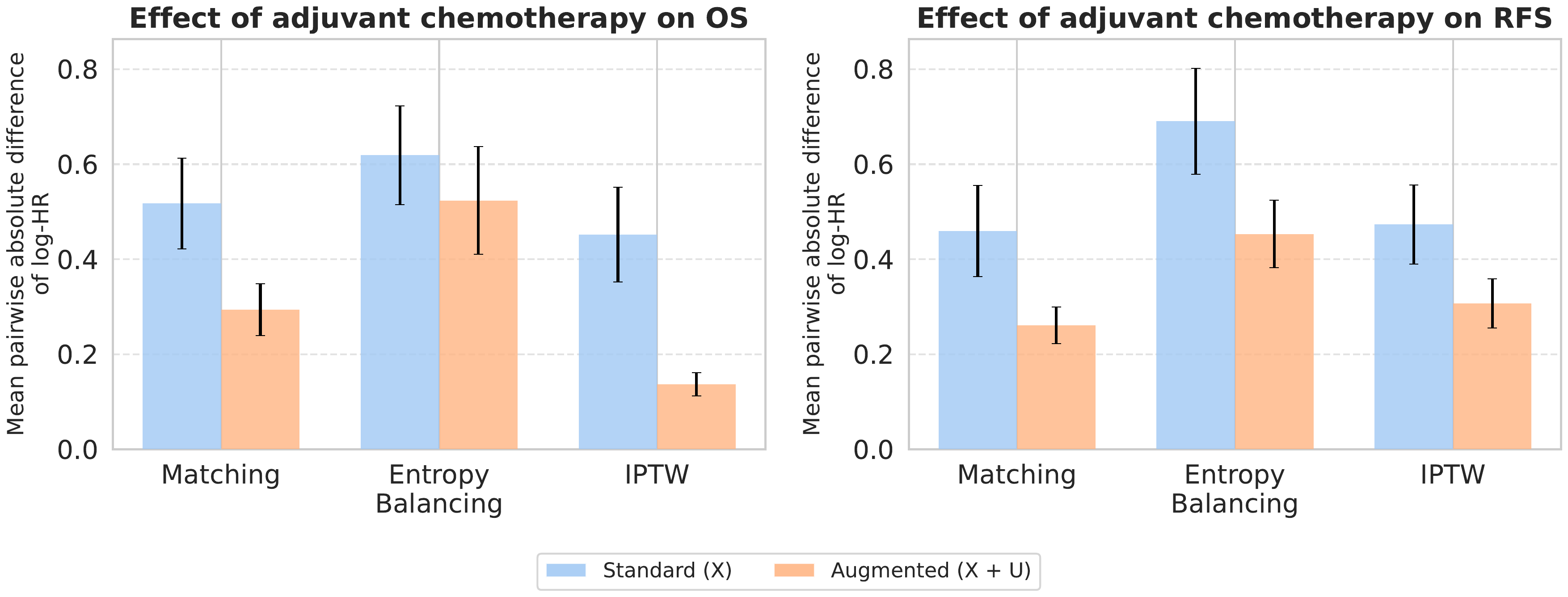}
\caption{Between-center heterogeneity in treatment-effect estimates for the CRLM multi-center cohort. Bars represent the mean pairwise absolute difference in log-hazard ratios (log-HR) across centers, averaged over all center pairs within each method and configuration. Results are shown for matching, entropy balancing, and inverse propensity weighting using either baseline covariates $X$ or the augmented representation $(X,\tilde{U})$. Error bars denote the standard error across pairwise comparisons.}
\label{fig:crlm_loghr_comp}
\end{figure*}

Across all balancing methods and both outcomes (recurrence and overall survival), incorporating $\tilde{U}$ reduces the mean absolute pairwise deviation of HR estimates across centers, as shown in Figure~\ref{fig:crlm_loghr_comp}. For a given balancing method $m$, variant $v$, and outcome $e$, this deviation is computed as

\begin{equation}
D_{m,v,e} =
\frac{1}{\binom{C}{2}}
\sum_{i<j}
\left|
\log(HR_{i,m,v,e}) - \log(HR_{j,m,v,e})
\right|,
\end{equation}

\noindent
where $C$ denotes the number of centers and $HR_{i,m,v,e}$ is the estimated hazard ratio for center $i$ under method $m$, variant $v$, and outcome $e$. In most settings, the reduction in $D_{m,v,e}$ after incorporating $\tilde{U}$ is statistically meaningful, indicating that adjusting for $\tilde{U}$ within each center reduces residual confounding and leads to more consistent treatment-effect estimates across centers.

\begin{table}[htbp]
\centering
\caption{Validation approach 3, first component: reduction in mean pairwise absolute deviation after adjusting for $X, \tilde{U}$ for 6 CRLM centers.}
\label{tab:crlm_comp}
\small
\begin{tabular}{lr}
\toprule
Statistic & Value \\
\midrule
Cells with improvement ($\Delta>0$) & 6 / 6 \\
Sign test $p$-value (one-sided) & 0.0156 \\
Wilcoxon signed-rank $p$-value (one-sided) & 0.0156 \\
Median cell-level $\Delta$ & 0.1343 \\
Mean cell-level $\Delta$ & 0.1467 \\
\bottomrule
\end{tabular}
\end{table}

The first component of Validation 3 (Table~\ref{tab:crlm_comp}) indicates that adjusting for both $X$ and $\tilde{U}$ reduces the average absolute pairwise differences in HR estimates across centers in all six evaluated cases. This reduction is statistically significant under both the sign test and the Wilcoxon signed-rank test. The differences in all pairs of datasets and balancing methods are reported in the Appendix Table \ref{tab:event_method_comparison}.

The second component of the cross-center validation, in contrast, examines differences in patient case mix across centers. In this setting, we construct pairwise comparisons by balancing center populations to each other through reweighting. We performed the statistical analysis described in Section~\ref{sec:validation_stats}, focusing on center pairs where adjustment using $X$ alone increased disagreement in adjusted 5-year survival relative to the crude estimates. Such cases suggest the presence of unobserved prognostic differences in patient populations across centers. We therefore examined whether incorporating $\tilde{U}$ in the adjustment could reduce these discrepancies.

This is precisely what we observe in Table~\ref{tab:u_validation_summary_combined}. Across both outcomes (overall survival and recurrence), and in both the full cohort and the surgery alone subset, adding $\tilde{U}$ reduced the survival gap in the majority of center pairs for which adjustment on $X$ alone had increased disagreement. Moreover, across several adjustment methods, these improvements occur at rates that are statistically significant under the corresponding binomial tests, supporting the interpretation that $\tilde{U}$ captures the unobserved prognostic component of cross-center differences in patient case mix, which in turn contributes to cross-center survival variation.

\begin{table*}[ht]
\centering
\caption{Validation 3, second component: Cross-center validation of the latent prognostic factor $\tilde{U}$. Among center pairs where adjustment for observed covariates $X$ increased disagreement in adjusted 5-year survival, we evaluated whether adding $\tilde{U}$ reduced the gap. Results are shown for the full cohort and for the surgery alone subset. The adjustment happens in the form of weights calculated to balance the pairs of centers with different balancing methods. These weights are then used in the Kaplan-Meier estimation.}
\label{tab:u_validation_summary_combined}
\footnotesize
\setlength{\tabcolsep}{4pt}
\resizebox{0.99\textwidth}{!}{
\begin{tabular}{lllcccc}
\toprule
\textbf{Subset} & \textbf{Event} & \textbf{Method} & \textbf{Pairs Evaluated} & \textbf{Pairs Widened by $X$} & \textbf{$U$ Fixed (\%)} & \textbf{Binomial $p$} \\
\midrule
Full cohort & Mortality & Matching & 30 & 19 (63.3\%) & 16/19 (84.2\%) & 0.0022 \\
Full cohort & Mortality & IPTW & 30 & 15 (50.0\%) & 13/15 (86.7\%) & 0.0037 \\
Full cohort & Mortality & Entropy Balancing & 16 & 10 (62.5\%) & 9/10 (90.0\%) & 0.0107 \\
Full cohort & Recurrence & Matching & 30 & 10 (33.3\%) & 8/10 (80.0\%) & 0.0547 \\
Full cohort & Recurrence & IPTW & 30 & 18 (60.0\%) & 15/18 (83.3\%) & 0.0038 \\
Full cohort & Recurrence & Entropy Balancing & 16 & 9 (56.2\%) & 7/9 (77.8\%) & 0.0898 \\
\midrule
Surgery alone & Mortality & Matching & 30 & 12 (40.0\%) & 9/12 (75.0\%) & 0.0730 \\
Surgery alone & Mortality & IPTW & 30 & 14 (46.7\%) & 14/14 (100.0\%) & 0.0001 \\
Surgery alone & Mortality & Entropy Balancing & 6 & 4 (66.7\%) & 4/4 (100.0\%) & 0.0625 \\
Surgery alone & Recurrence & Matching & 30 & 14 (46.7\%) & 11/14 (78.6\%) & 0.0287 \\
Surgery alone & Recurrence & IPTW & 30 & 17 (56.7\%) & 14/17 (82.4\%) & 0.0064 \\
Surgery alone & Recurrence & Entropy Balancing & 6 & 5 (83.3\%) & 5/5 (100.0\%) & 0.0312 \\
\bottomrule
\end{tabular}
}
\end{table*}

\section{Discussion}

\indent In this study, we developed and evaluated a framework to address unobserved confounding in observational survival analyses by inferring and balancing a latent prognostic factor, $\tilde{U}$. We examined the framework across three complementary validation settings: comparison of observational HR estimates against benchmark randomized trial effects, application to randomized datasets as a negative-control setting, and multicenter colorectal liver metastasis analyses evaluating whether balancing $\tilde{U}$ reduces between-center heterogeneity in treatment effects and patient case mix. Taken together, these findings suggest that $\tilde{U}$ captures clinically meaningful latent prognostic information not contained in the recorded covariates alone.

\indent The first validation strategy, comparison with benchmark randomized trials, addressed the central question of whether balancing $\tilde{U}$ improves treatment-effect estimation in confounded observational data. Across the observational cohorts studied here, this was generally the case. Importantly, improved agreement with the benchmark trial should not be interpreted as implying that observational estimates ought to become identical to the randomized estimate. Even with better control of confounding, hazard ratios may still differ because they depend not only on confounding, but also on differences in eligibility criteria, treatment definitions, follow-up duration, and other features distinguishing real-world cohorts from trial populations. For example, in RPS, the benchmark STRASS trial used abdominal recurrence-free survival, a trial-specific endpoint that cannot be directly recreated in real-world datasets, where the available outcome is local recurrence, a similar but not identical endpoint. The relevant finding is therefore not exact numerical convergence, but a systematic shift toward more plausible estimates when an external benchmark is available.

\indent The second validation strategy, application to randomized datasets, provides an important complementary check. If $\tilde{U}$ reflects latent prognosis rather than arbitrary statistical noise, it should be balanced across treatment arms under randomization and should not materially distort treatment-effect estimates when added to the analysis. This is what we observed. In that sense, the randomized analyses function as a negative control: they support the interpretation of $\tilde{U}$ as a prognostic construct while arguing against the possibility that its apparent utility in observational cohorts simply reflects overfitting.

\indent The multicenter CRLM analyses extend the evaluation beyond comparison with an external randomized benchmark and instead test whether $\tilde{U}$ helps explain two distinct forms of between-center heterogeneity. The first concerns hidden bias in who receives adjuvant chemotherapy within a center. If treatment allocation remains influenced by latent prognosis after adjustment for recorded covariates, then estimated chemotherapy effects may differ across centers partly for confounding-related reasons rather than because the true treatment effect is fundamentally different. The observed reduction in between-center dispersion of chemotherapy hazard-ratio estimates after balancing $\tilde{U}$ suggests that part of the latent structure underlying chemotherapy selection was shared across cohorts and was captured by $\tilde{U}$. This is biologically and clinically plausible, because many of the factors that influence receipt of chemotherapy in routine practice—such as frailty, limited tolerance for systemic treatment, occult disease burden, and other incompletely measured aspects of patient fitness—are not specific to any one center but recur across cohorts. Even so, we did not expect identical hazard ratios across centers, because treatment itself was not fully standardized across cohorts, with differences in chemotherapy regimens, biologic agents, treatment sequencing, overall patient selection, and follow-up.

\indent The second multicenter question concerns hidden differences in patient case mix across centers. Here, the aim was not to reduce confounding in treated-versus-untreated comparisons within a cohort, but to determine whether latent differences in prognosis between center populations contributed to cross-center survival differences beyond those explained by recorded covariates. In several center pairs, balancing on measured covariates alone widened the survival gap, indicating that observed case mix had partly masked underlying outcome differences. Subsequent incorporation of $\tilde{U}$ narrowed these residual gaps in many settings, supporting the interpretation that $\tilde{U}$ captures the unobserved prognostic component of cross-center differences in patient case mix. At the same time, these differences were not expected to disappear entirely. After balancing on recorded covariates and $\tilde{U}$, any remaining survival differences are more plausibly attributable to residual case-mix differences not captured by the framework or to true center-level differences in care including differences in surgical quality, perioperative management, surveillance, and follow-up.

\indent An additional strength of the framework is that its performance was not restricted to a single analytic specification. We evaluated it across multiple balancing strategies and prespecified hyperparameter ranges, and summarized performance across parameter combinations rather than selecting a single favorable run. This reduces the likelihood that the apparent benefit of $\tilde{U}$-augmented analyses was driven by post hoc tuning. The framework also remained informative across different prognostic model inputs, including externally developed prognostic models, suggesting that its usefulness does not depend on a single modeling pipeline. This is particularly important because, in medical cohorts and especially in surgical oncology cohorts, prognostic models often achieve only modest discrimination. Retroperitoneal sarcoma and colorectal liver metastases are good examples of settings in which few published models achieve a c-index above 0.7.\cite{kokkinakis2024clinical, tan2016histology, sasaki2021performance} Demonstrating that the framework can still perform well in such settings is important for its broader applicability. At the same time, the scope of the framework should be stated clearly. It was designed to address residual confounding arising from latent prognostic heterogeneity, not to resolve all sources of bias in observational treatment comparisons. In particular, it was not explicitly designed to correct immortal time bias  which may be relevant in comparisons such as surgery alone versus surgery plus adjuvant therapy.  Such bias can arise because patients who go on to receive adjuvant treatment must, by definition, survive event-free long enough after surgery to initiate therapy, thereby creating a potential artificial advantage if outcomes are compared from the date of surgery. Accordingly, our findings should not be interpreted as showing that $\tilde{U}$ augmentation corrects immortal time bias per se. Rather, they suggest that the framework can remain informative in real-world settings where multiple sources of bias coexist. In applications where immortal time bias is likely to be important, the framework should ideally be combined with methods developed specifically for that problem, such as landmark or time-dependent analyses.  

\indent   The disease-specific patterns observed here also help clarify the settings in which the framework is most likely to add value. In GIST, conventional multivariable adjustment already shifted estimates toward the benchmark trial effect, consistent with the strong prognostic and treatment-selection relevance of routinely recorded variables such as tumor size, mitotic index, and tumor site. In contrast, in RPS, latent-factor augmentation appeared more informative, likely because prognostic models are weaker and radiotherapy allocation may depend more heavily on unrecorded clinical judgment. Indeed, both the published analysis of the RPS dataset used in our experiments and our own findings showed that multivariable regression produced little change relative to the univariable estimate, suggesting that the commonly recorded prognostic factors were not the principal drivers of perioperative radiotherapy use.\cite{kelly2015comparison} Rather, selection for RT may have been influenced more strongly by unmeasured factors reflecting radiation oncologists’ clinical judgment, including subtler aspects of disease presentation not captured in the available covariates. Taken together, these findings suggest that the framework may be especially valuable in clinical settings where substantial residual prognostic heterogeneity persists despite adjustment for standard recorded covariates.  

\indent The implications of this work extend beyond estimation of average treatment effects. If observational cohorts can be made more comparable not only with respect to measured covariates but also with respect to latent prognostic imbalance, they may provide a stronger basis for downstream analyses aimed at identifying subgroups with differential treatment benefit.  This is particularly important because real-world datasets are typically much larger than RCTs, which are costly and time-intensive, and may better reflect contemporary practice patterns and patient selection. However, the intention of the proposed framework is not to replace randomized trials—either for estimating average treatment effects or for discovering subgroup-specific effects—but to supplement them in settings where trials are infeasible, underpowered for subgroup analyses, or not fully representative of routine care. If confirmed in independent external datasets, this approach may support more credible treatment-effect estimation and more reliable downstream precision-oncology applications.


\section*{Declaration of interests}

The authors declare no competing interests.

\section*{Data sharing}

Patient data used in this study were obtained from participating institutions under data use agreements and cannot be publicly shared due to privacy restrictions. 

\section*{Acknowledgments}

The authors would like to thank Dr. Nefeli Bampatsikou for the excellent technical support and Ms. Eftychia Ioustine Margonis for her support. This research was supported by the NCI Cancer Center Support Grant P30 CA008748 (G.A.M.) and the Abdul Latif Jameel Clinic for Machine Learning in Health (D.B.).

\bibliography{ref}

\appendix
\onecolumn
\renewcommand{\thefigure}{\thesection.\arabic{figure}}
\renewcommand{\thetable}{\thesection.\arabic{table}}
\renewcommand{\theequation}{\thesection.\arabic{equation}}

\setcounter{figure}{0}
\setcounter{table}{0}
\setcounter{equation}{0}

\renewcommand{\thesection}{\arabic{section}}

\section{Sensitivity Analysis}\label{app_sect:sens_analysis}

All sensitivity plots report the mean absolute log-HR error relative to the benchmark hazard ratio, averaged across hyperparameter configurations, with standard errors shown as error bars.

\begin{figure}[htbp!]
\centering
\includegraphics[width=0.9\linewidth]{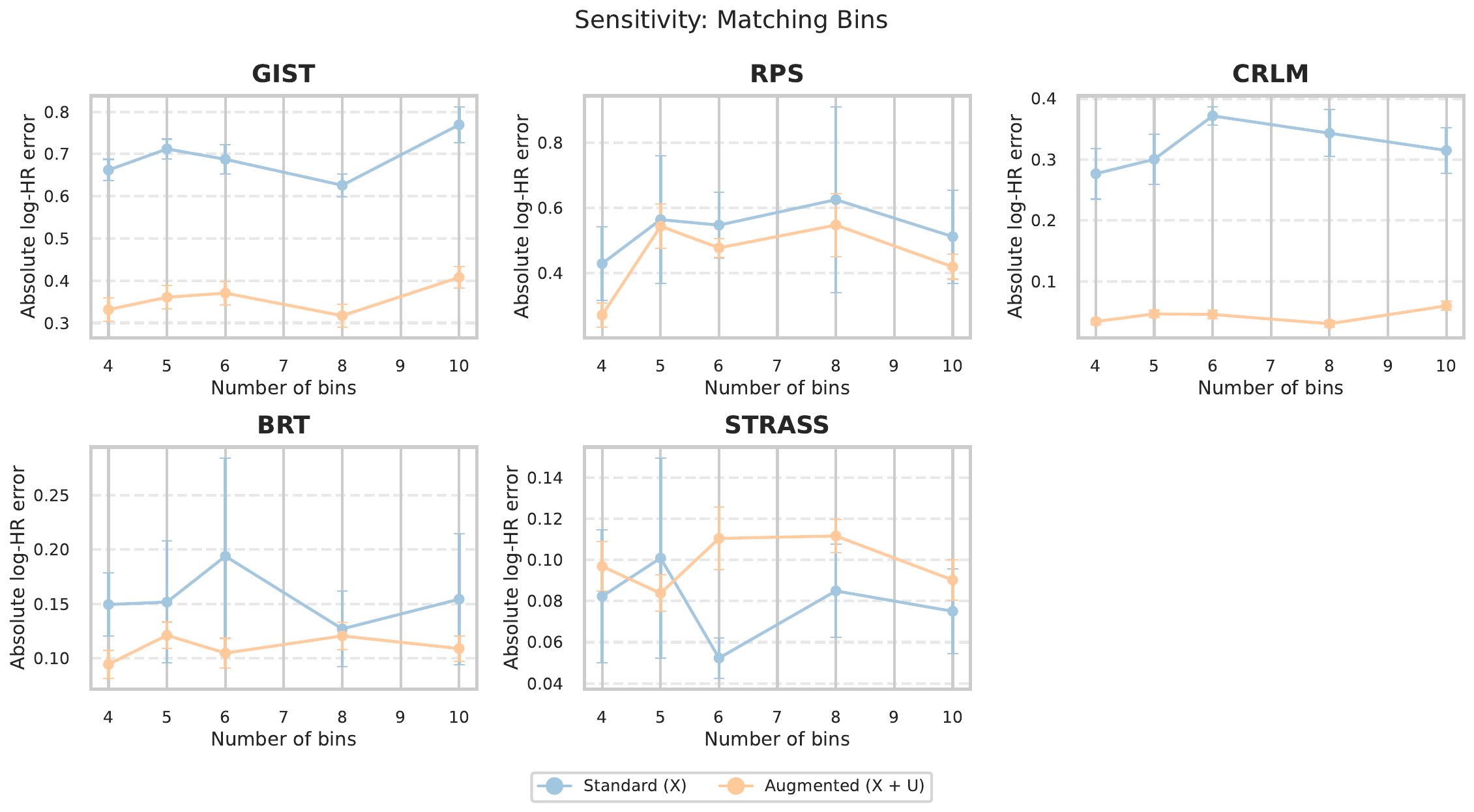}
\caption{Sensitivity of treatment-effect estimation to the number of bins used in prognostic matching for the observational and RCT cohorts. The plotted values represent the mean absolute log-HR error relative to the benchmark hazard ratio, averaged across hyperparameter configurations. Results are shown with and without the latent factor $\tilde{U}$. Error bars denote the standard error across configurations.}
\label{app_fig:sens_bins_obs}
\end{figure}

\begin{figure}[htbp!]
\centering
\includegraphics[width=0.9\linewidth]{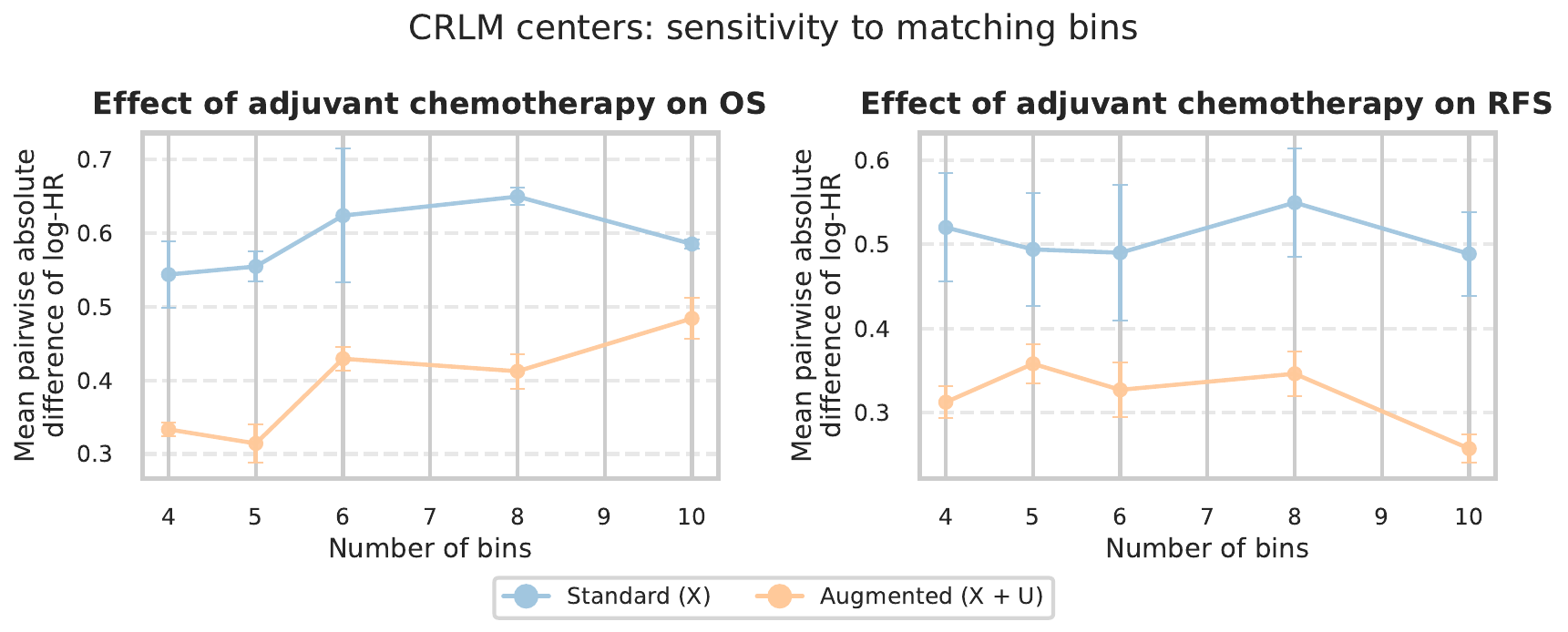}
\caption{Sensitivity of treatment-effect estimation to the number of bins used in prognostic matching for the 6 CRLM centers. The plotted values represent the mean pairwise absolute difference of log-HR between centers, averaged across hyperparameter configurations. Results are shown with and without the latent factor $\tilde{U}$. Error bars denote the standard error across configurations.}
\label{app_fig:sens_bins_crlm}
\end{figure}

\begin{figure}[htbp!]
\centering
\includegraphics[width=0.9\linewidth]{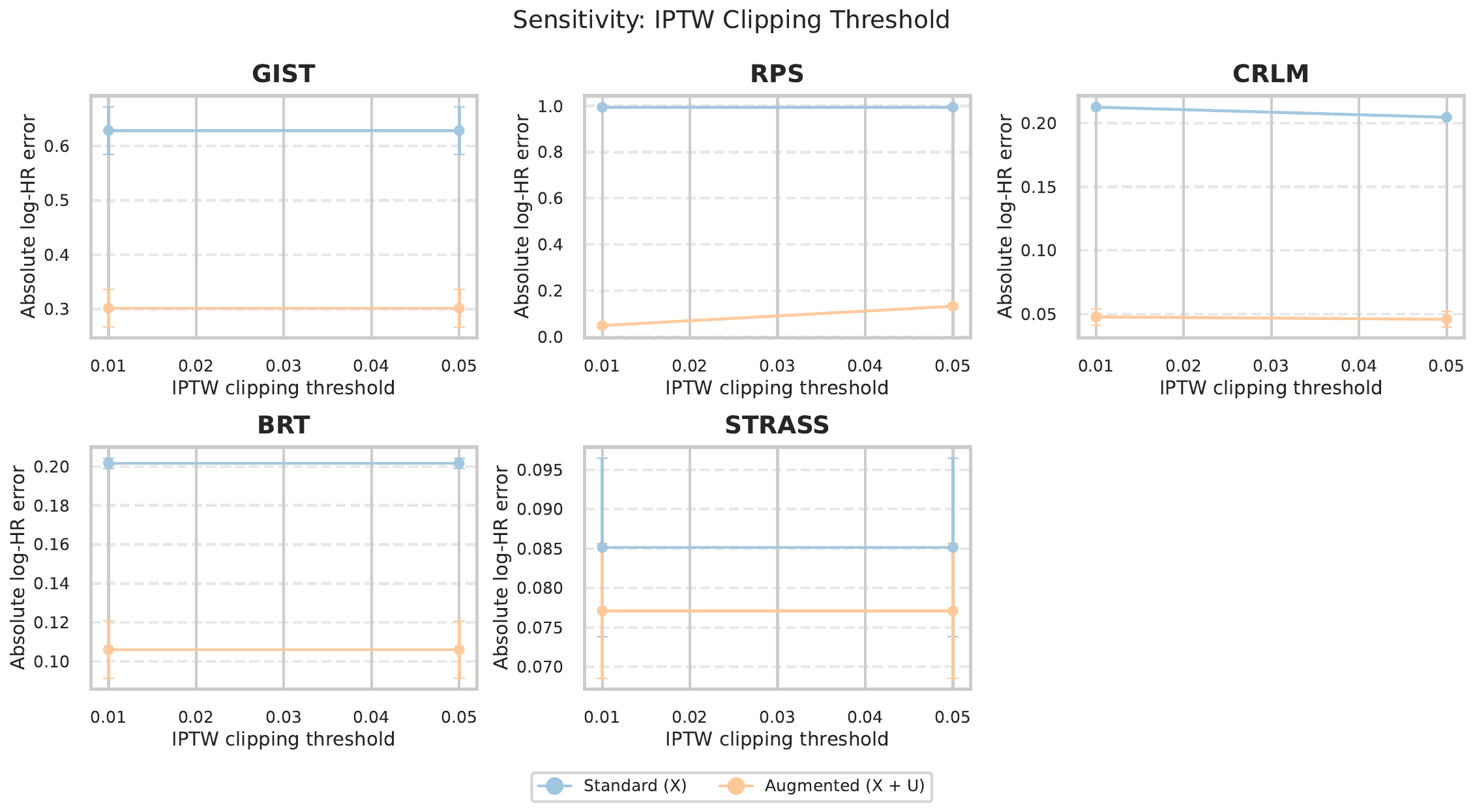}
\caption{Sensitivity of inverse propensity weighting (IPTW) to the clipping threshold applied to the estimated propensity scores for the observational and RCT cohorts. The plotted values represent the mean absolute log-HR error relative to the benchmark hazard ratio across hyperparameter configurations. Results are shown for models using baseline covariates $X$ and the augmented representation $(X,\tilde{U})$. Error bars denote the standard error across configurations.}
\label{app_fig:sens_clip_obs}
\end{figure}

\begin{figure}[htbp!]
\centering
\includegraphics[width=0.9\linewidth]{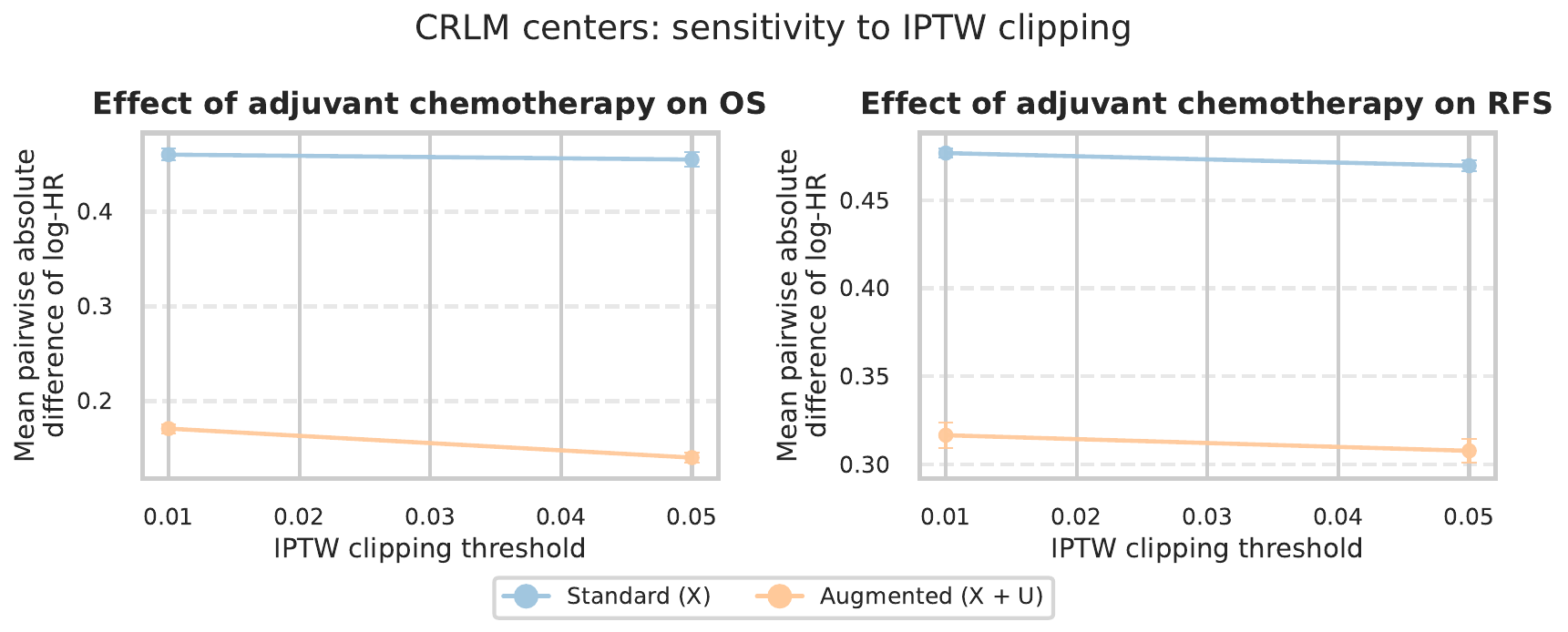}
\caption{Sensitivity of inverse propensity weighting (IPTW) to the clipping threshold applied to the estimated propensity scores for the 6 CRLM centers. The plotted values represent the mean pairwise absolute difference of log-HR between centers, averaged across hyperparameter configurations. Results are shown for models using baseline covariates $X$ and the augmented representation $(X,\tilde{U})$. Error bars denote the standard error across configurations.}
\label{app_fig:sens_clip_crlm}
\end{figure}

\begin{figure}[htbp!]
\centering
\includegraphics[width=0.9\linewidth]{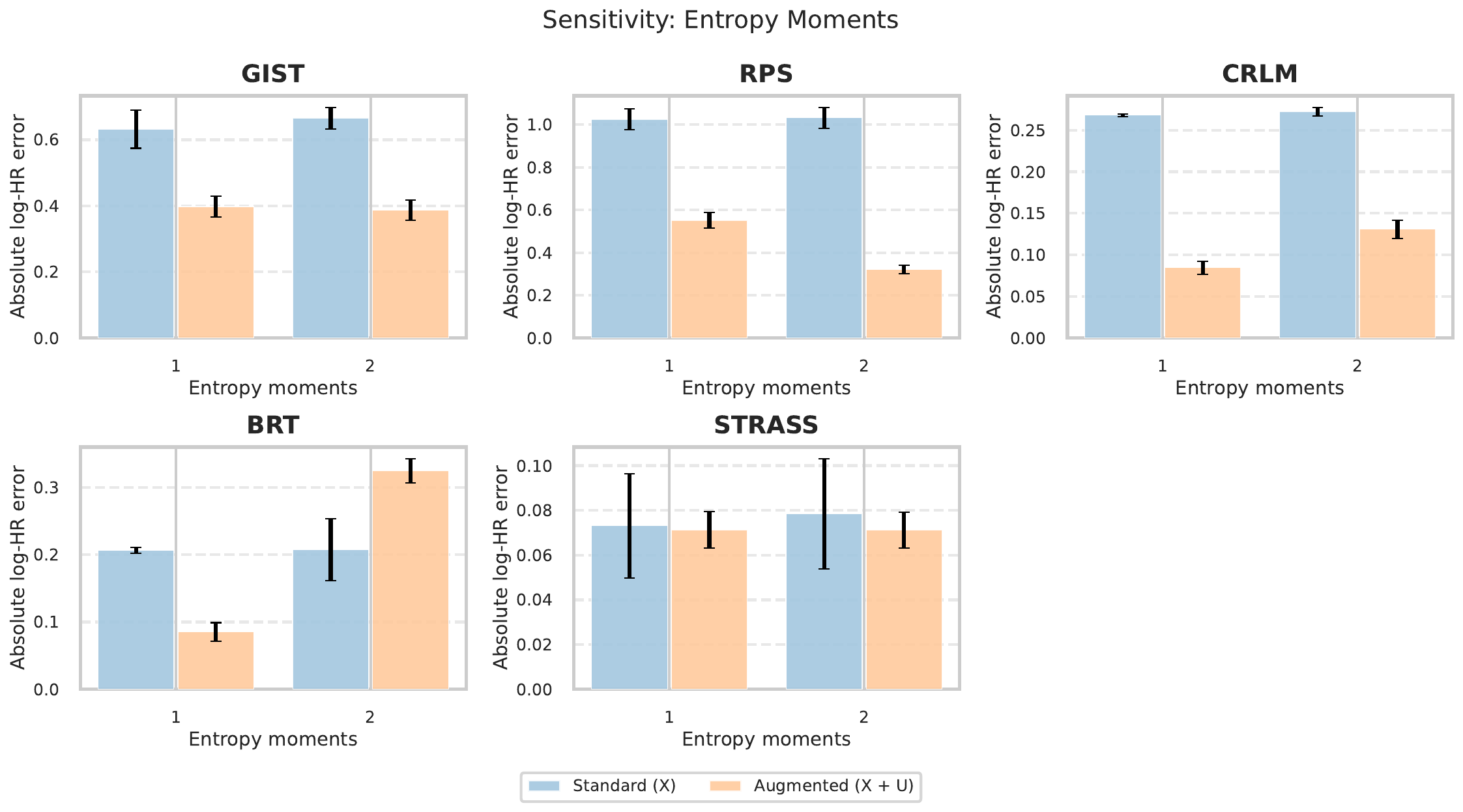}
\caption{Sensitivity of entropy balancing to the number of balanced moments for the observational and RCT datasets. Bars represent the mean absolute log-HR error relative to the benchmark hazard ratio, averaged across hyperparameter configurations. Results compare models using baseline covariates $X$ with those augmented by the latent factor $\tilde{U}$. Error bars denote the standard error across configurations.}
\label{app_fig:sens_moments_obs}
\end{figure}

\begin{figure}[htbp!]
\centering
\includegraphics[width=0.9\linewidth]{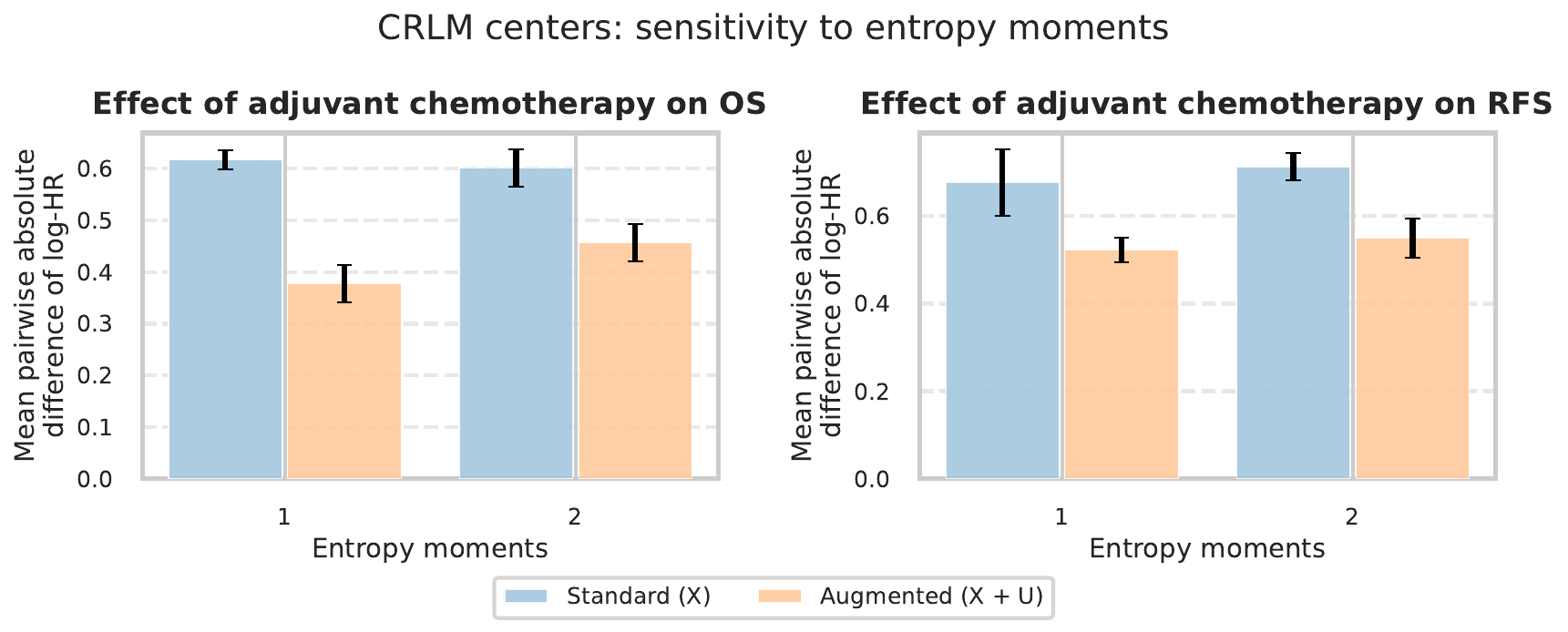}
\caption{Sensitivity of entropy balancing to the number of balanced moments for the 6 CRLM centers. Bars represent the mean pairwise absolute difference of log-HR between centers, averaged across hyperparameter configurations. Results compare models using baseline covariates $X$ with those augmented by the latent factor $\tilde{U}$. Error bars denote the standard error across configurations.}
\label{app_fig:sens_moments_crlm}
\end{figure}

\begin{figure}[htbp!]
\centering
\includegraphics[width=0.9\linewidth]{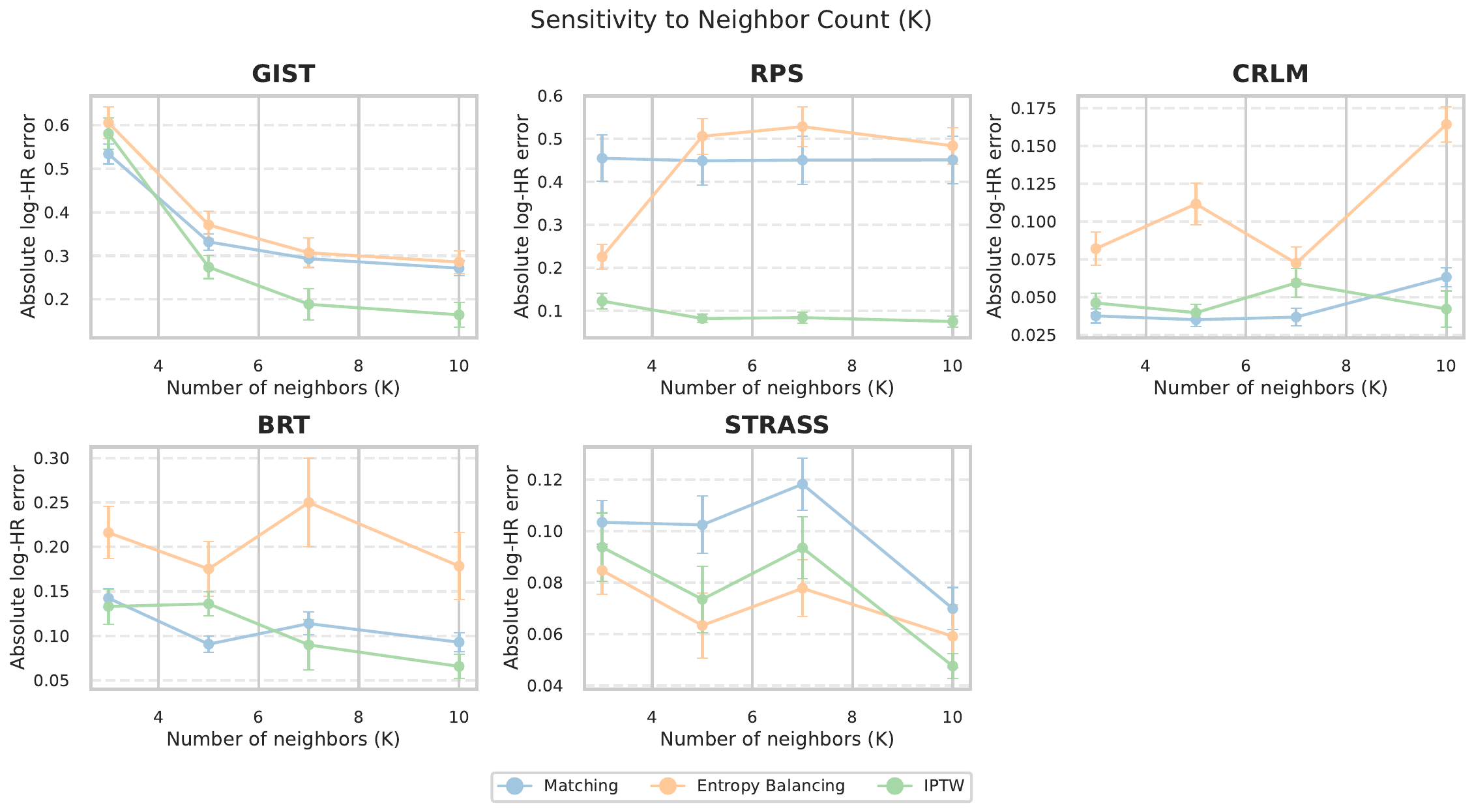}
\caption{Sensitivity of the framework to the number of nearest neighbors ($K$) used when constructing the latent prognostic factor $\tilde{U}$ for the observational and RCT datasets. Lines represent the mean absolute log-HR error relative to the benchmark hazard ratio across configurations for each balancing method. Error bars denote the standard error across configurations.}
\label{app_fig:sens_K_obs}
\end{figure}

\begin{figure}[htbp!]
\centering
\includegraphics[width=0.9\linewidth]{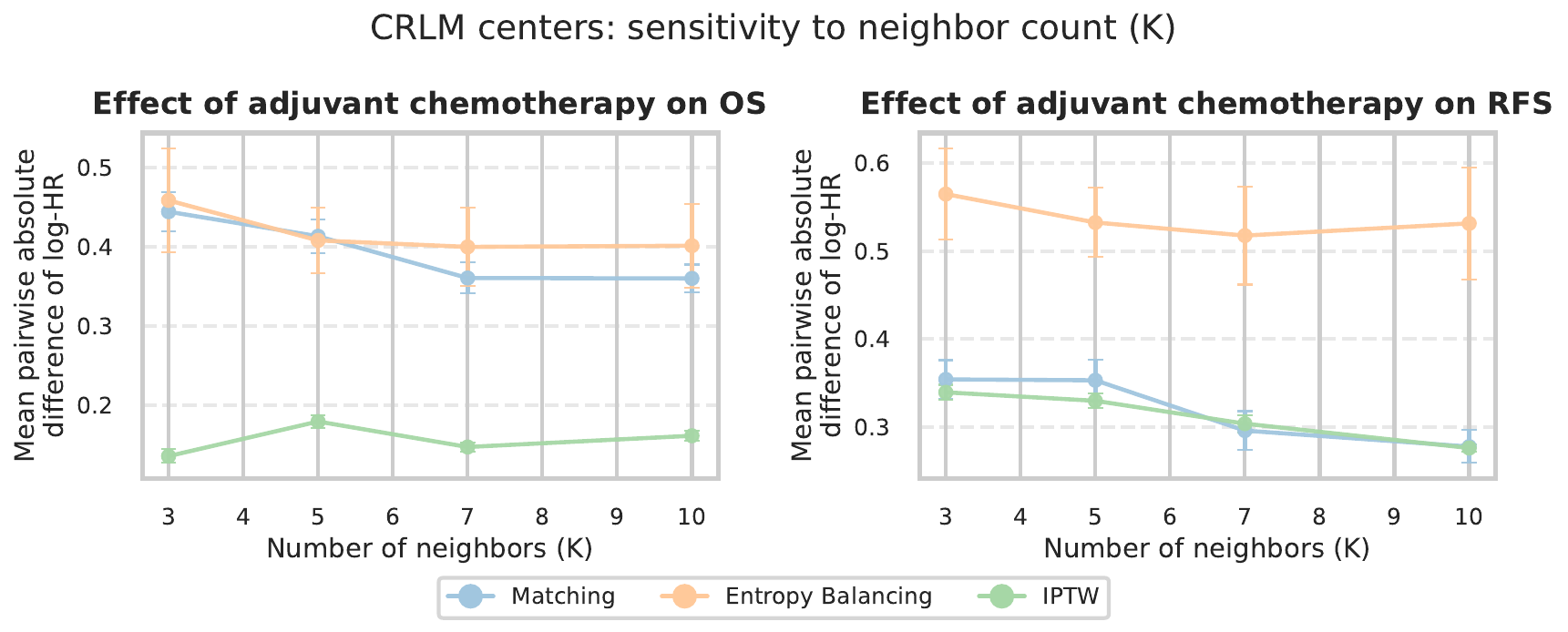}
\caption{Sensitivity of the framework to the number of nearest neighbors ($K$) used when constructing the latent prognostic factor $\tilde{U}$ for the 6 CRLM centers. Lines represent the the mean pairwise absolute difference of log-HR between centers, averaged across hyperparameter configurations for each balancing method. Error bars denote the standard error across configurations.}
\label{app_fig:sens_K_crlm}
\end{figure}

\begin{figure}[htbp!]
\centering
\includegraphics[width=0.9\linewidth]{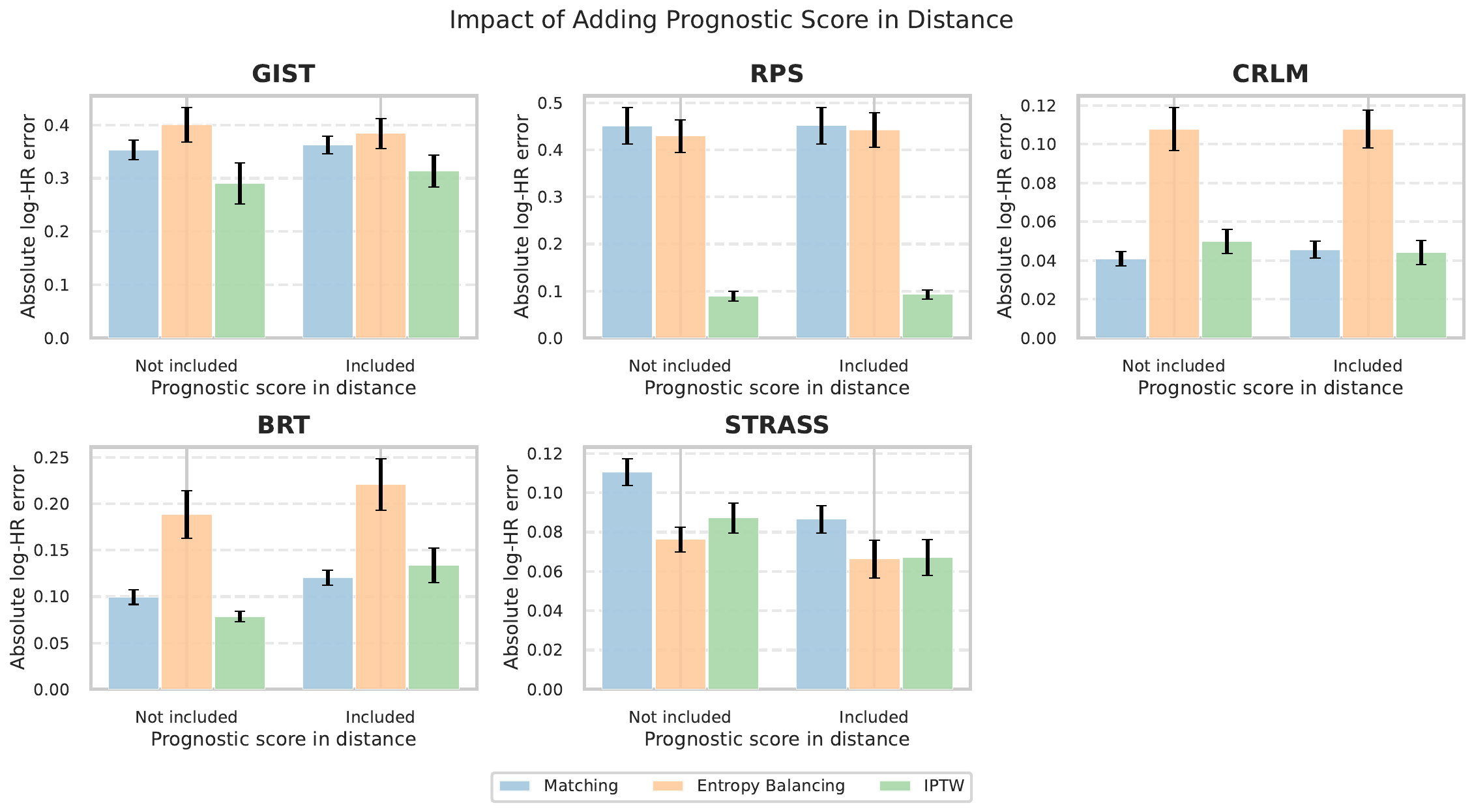}
\caption{Impact of incorporating the prognostic score in the distance metric used to construct $\tilde{U}$ for the observational and RCT datasets. Bars represent the mean absolute log-HR error relative to the benchmark hazard ratio across hyperparameter configurations for each balancing method. Error bars denote the standard error across configurations.}
\label{app_fig:sens_prog_obs}
\end{figure}

\begin{figure}[htbp!]
\centering
\includegraphics[width=0.9\linewidth]{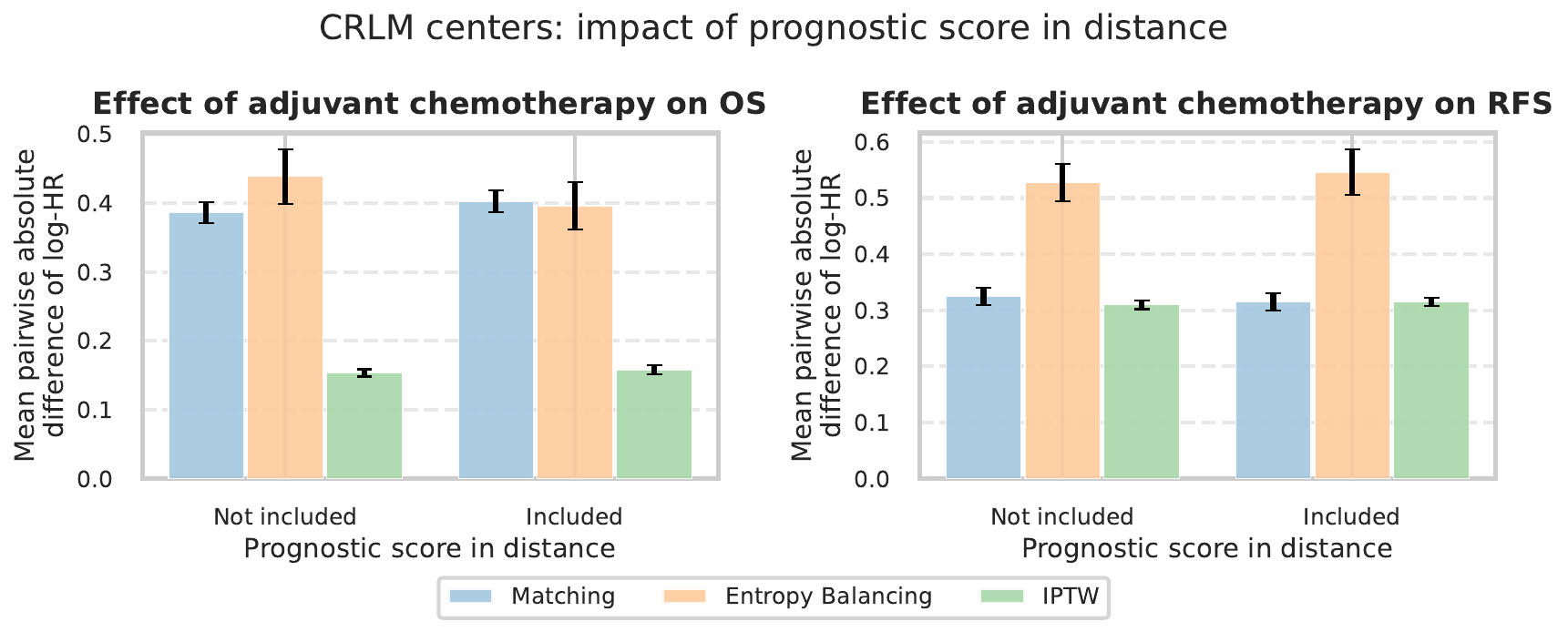}
\caption{Impact of incorporating the prognostic score in the distance metric used to construct $\tilde{U}$ for the 6 CRLM centers. Bars represent the mean pairwise absolute difference of log-HR between centers, averaged across hyperparameter configurations for each balancing method. Error bars denote the standard error across configurations.}
\label{app_fig:sens_prog_crlm}
\end{figure}

\begin{figure}[htbp!]
\centering
\includegraphics[width=0.9\linewidth]{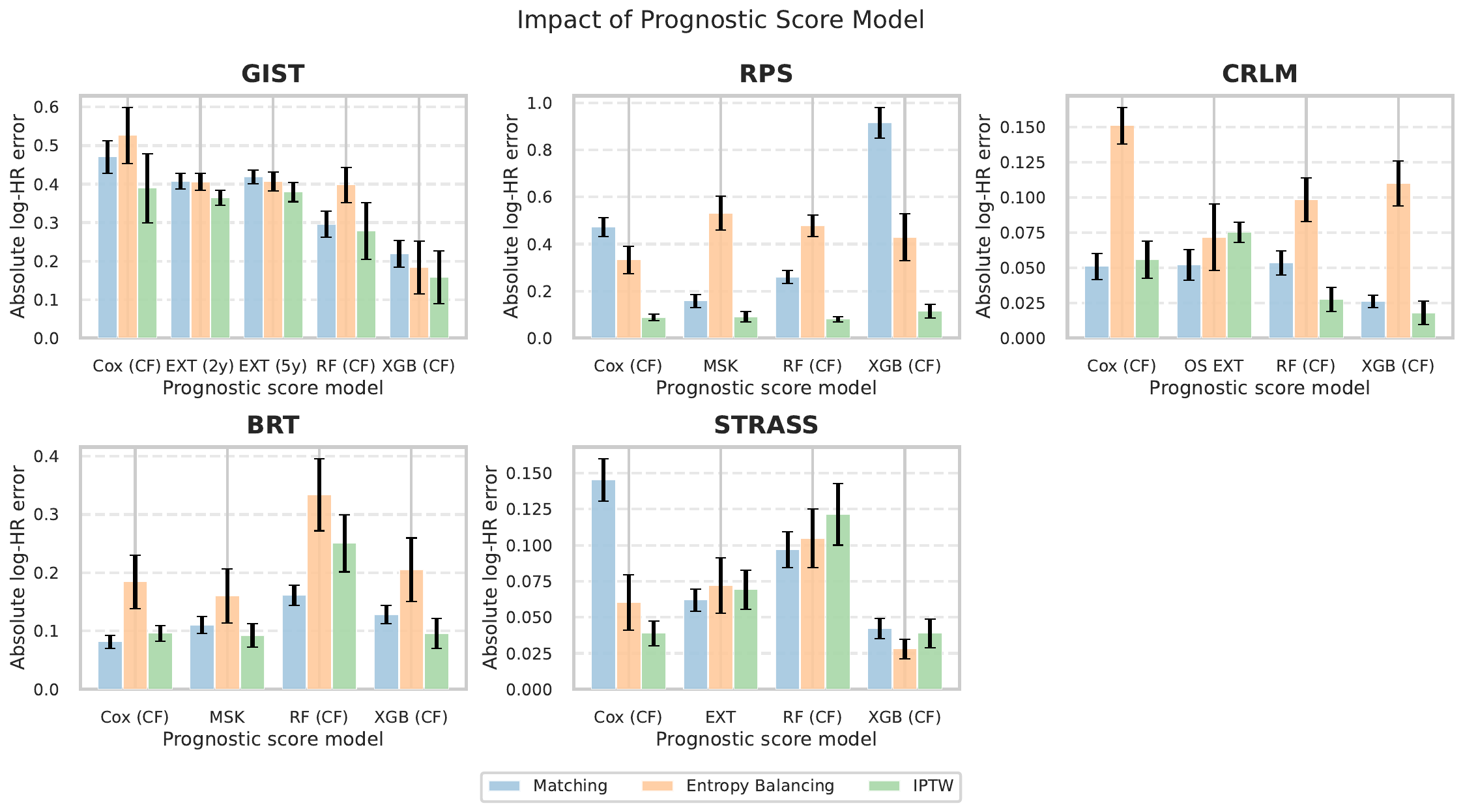}
\caption{Sensitivity of results to the choice of prognostic score model used to estimate the baseline risk component for the observational and RCT datasets. Bars represent the mean absolute log-HR error relative to the benchmark hazard ratio across hyperparameter configurations for each balancing method. Error bars denote the standard error across configurations.}
\label{app_fig:sens_score_model_obs}
\end{figure}

\begin{figure}[htbp!]
\centering
\includegraphics[width=0.9\linewidth]{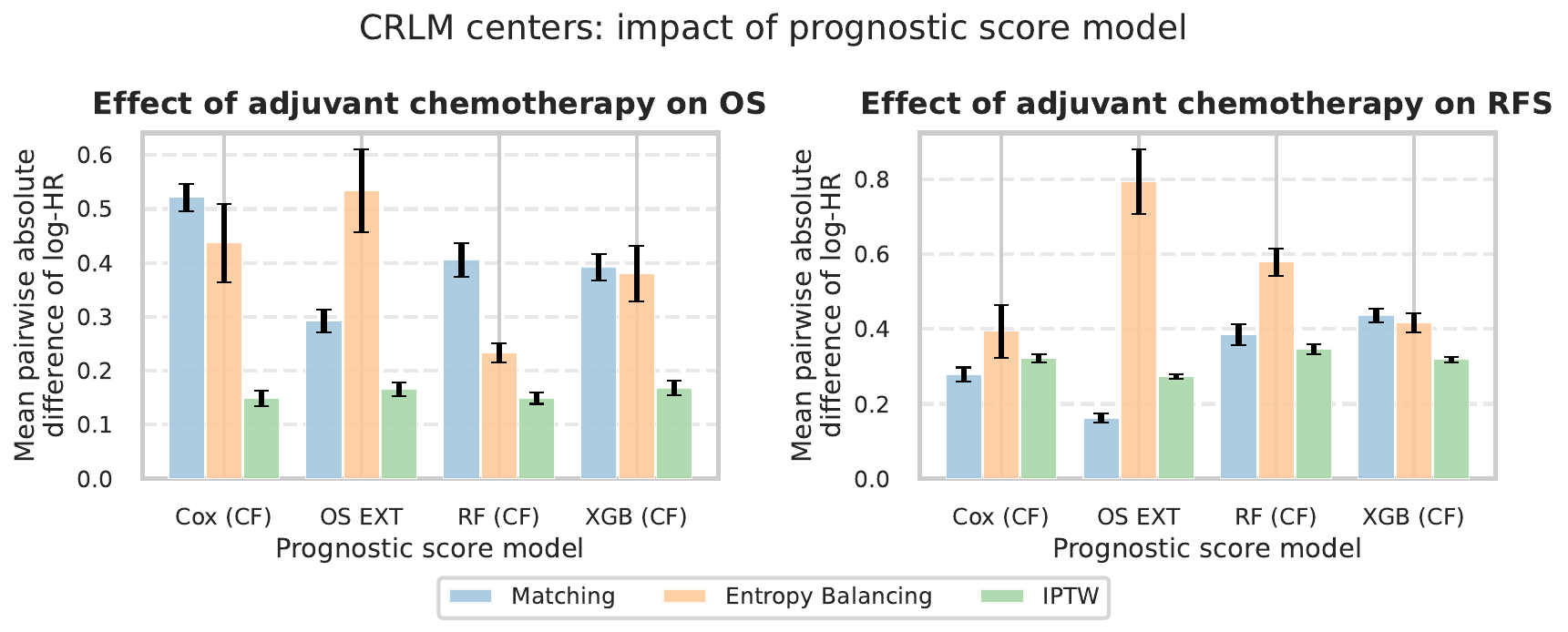}
\caption{Sensitivity of results to the choice of prognostic score model used to estimate the baseline risk component for the 6 CRLM centers. Bars represent the mean pairwise absolute difference of log-HR between centers, averaged across hyperparameter configurations for each balancing method. Error bars denote the standard error across configurations.}
\label{app_fig:sens_score_model_crlm}
\end{figure}

\section{Permutation and Ablation Experiments}

\begin{figure}[htbp!]
\centering
\includegraphics[width=0.9\linewidth]{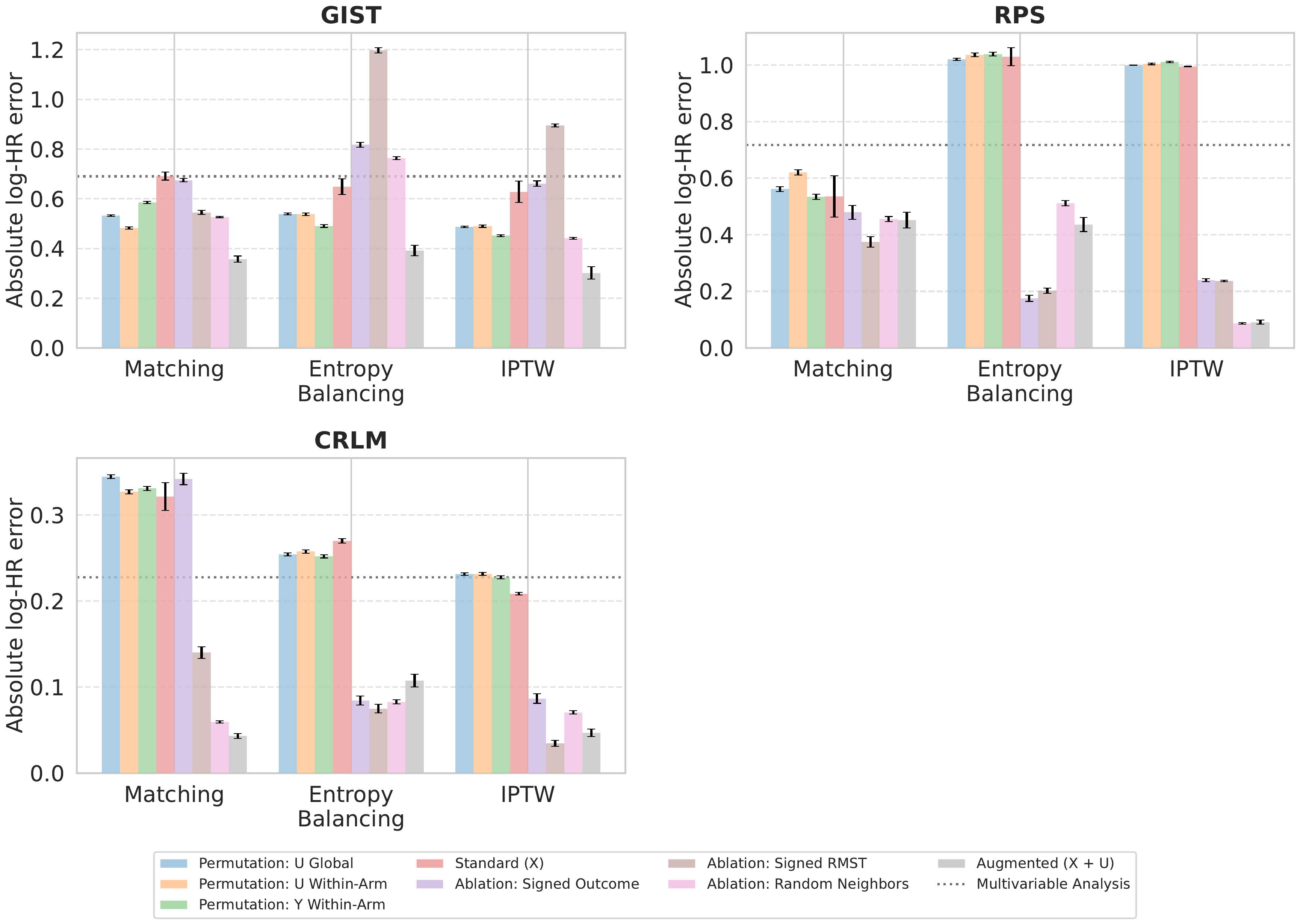}
\caption{Permutation and ablation test for the observational datasets. The latent factor $\tilde{U}$ is randomly permuted across patients before balancing. Bars represent the mean absolute log-HR error relative to the benchmark hazard ratio. Performance degradation relative to the non-permuted case indicates that the predictive signal of $\tilde{U}$ contributes to improved treatment-effect estimation. Error bars denote the standard error across configurations.}
\label{app_fig:obs_perm}
\end{figure}

\begin{figure}[htbp!]
\centering
\includegraphics[width=0.9\linewidth]{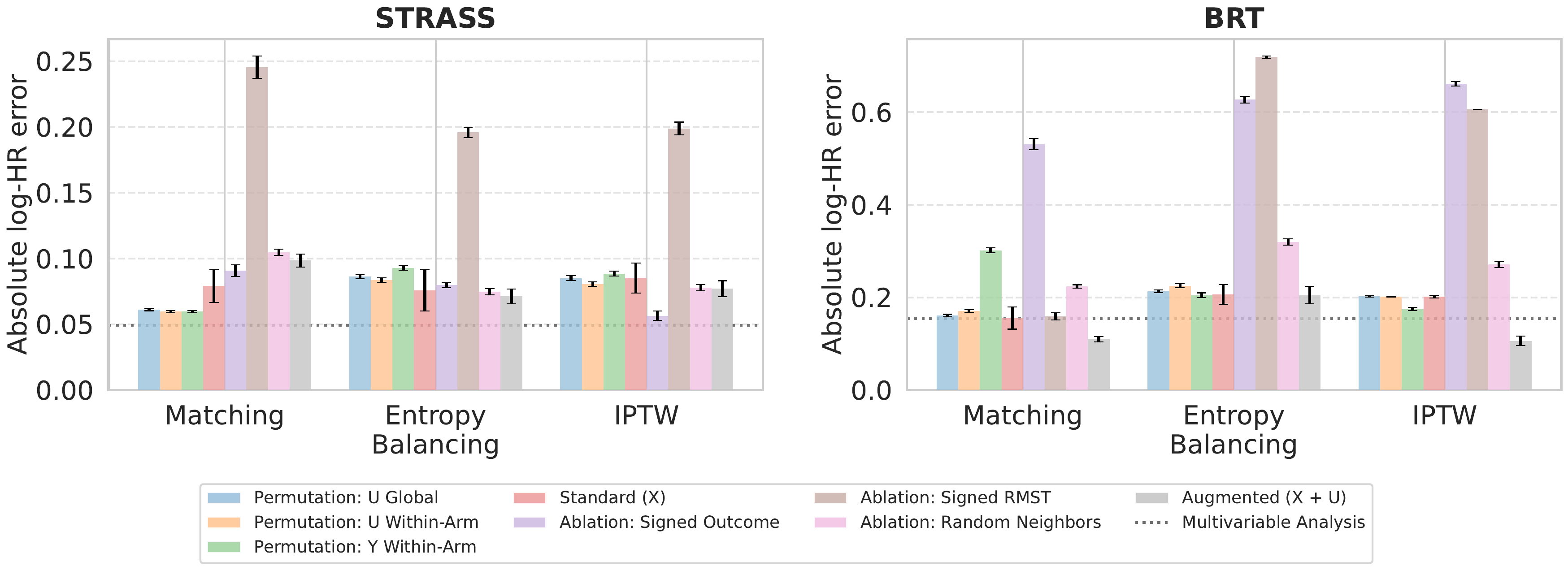}
\caption{Permutation and ablation test for the RCT datasets. The latent factor $\tilde{U}$ is randomly permuted across patients before balancing. Because treatment assignment is randomized, permuting $\tilde{U}$ does not materially change treatment-effect estimates, providing a negative control for the method. Error bars denote the standard error across configurations.}
\label{app_fig:rct_perm}
\end{figure}

\begin{figure}[htbp!]
\centering
\includegraphics[width=0.9\linewidth]{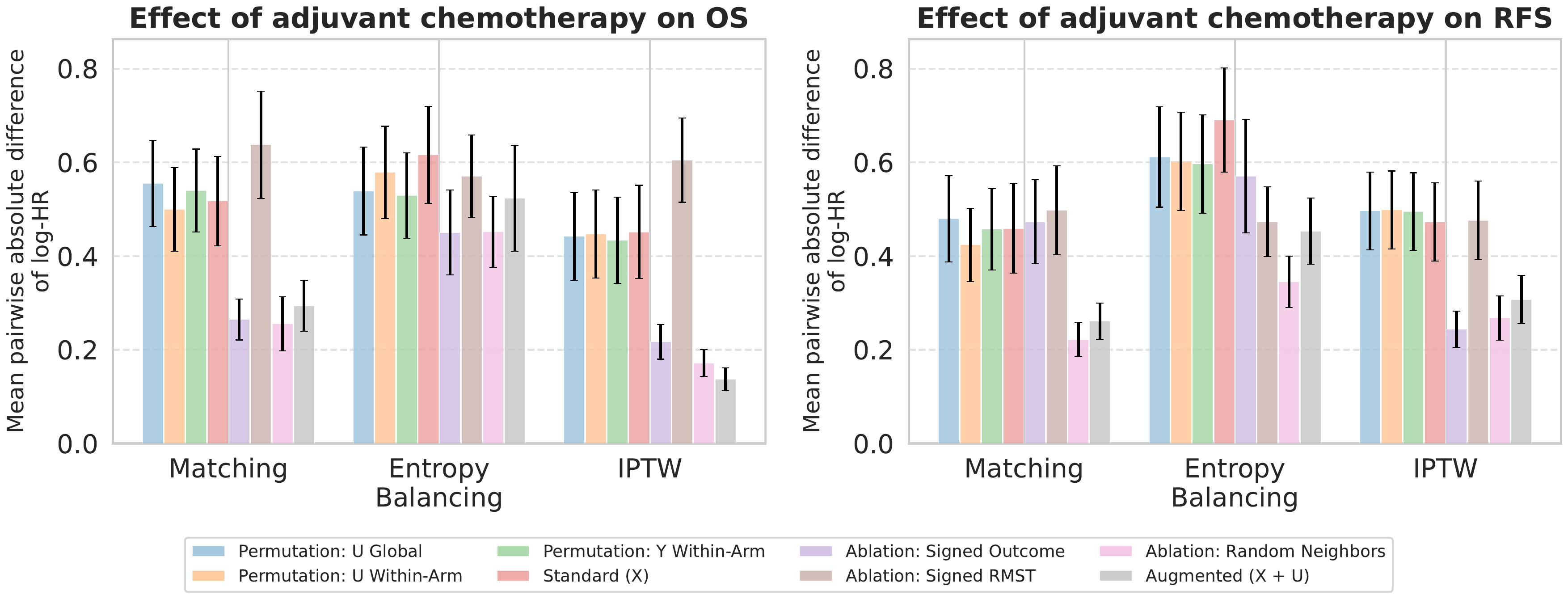}
\caption{Permutation test for the CRLM multi-center analysis. The latent factor $\tilde{U}$ is randomly permuted across patients prior to balancing. Bars represent the mean absolute pairwise HR difference across centers. The increase in cross-center variability after permutation indicates that the estimated $\tilde{U}$ captures meaningful latent prognostic structure shared across cohorts. Error bars denote the standard error across center pairs.}
\label{app_fig:crlm_comp_perm}
\end{figure}

\subsection{Permutation Experiments}

To examine whether the improvement observed after incorporating the latent factor $\tilde{U}$ could arise from mechanical artifacts rather than genuine latent structure, we conducted a series of permutation experiments.

\paragraph{Permutation of $Y$ (RMST) within treatment arms}

In the first experiment, we permuted the pseudo-RMST outcome values $Y$ within each treatment arm prior to computing $\tilde{U}$, while keeping the original (true) outcomes when estimating the final hazard ratios (HRs).

This manipulation breaks the individual-level association between $X$ and $Y$ within each arm, and consequently between $Y$ and $\tilde{U}$ during the construction of $\tilde{U}$. Importantly, however, it preserves the marginal distribution of $Y \mid T$ within each arm. In other words, the arm-level survival distribution remains intact, but the patient-specific mapping $X_i \leftrightarrow Y_i$ is destroyed.

If the method continued to produce improved HR estimates under this manipulation, it would suggest that the observed correction arises merely from injecting outcome-derived quantities into the balancing step (a mechanical artifact) rather than from $\tilde{U}$ capturing meaningful latent structure tied to the observed covariates. 

Instead, the experiment fails: the improvement disappears. This indicates that the specific pairing between $X_i$ and $Y_i$ is necessary for $\tilde{U}$ to encode meaningful information. The latent factor is not functioning as a generic outcome-based adjustment; it requires the genuine individual-level relationship between covariates and survival to operate properly.

\paragraph{Permutation of $\tilde{U}$ within treatment arms}

In the second experiment, we permuted $\tilde{U}$ within each treatment arm after its construction. This preserves the distribution of $\tilde{U}$ within each arm but breaks the individual-level mappings $\tilde{U}_i \leftrightarrow X_i$ and $\tilde{U}_i \leftrightarrow Y_i$.

A deterioration in results relative to the original method indicates that individual-level alignment matters: it is not sufficient for $\tilde{U}$ to encode only arm-level imbalance. Rather, the patient-specific latent information contained in $\tilde{U}_i$ contributes materially to the balancing step. The observed degradation supports the interpretation that $\tilde{U}$ carries individualized latent risk information rather than merely reflecting group averages.

\paragraph{Global permutation of $\tilde{U}$}

In the third experiment, we permuted $\tilde{U}$ across the entire dataset, ignoring treatment arms. This destroys not only the individual-level relationships but also the arm-level distribution of $\tilde{U}$.

Under this manipulation, $\tilde{U}$ becomes unrelated to treatment assignment and loses any confounding-relevant structure. As expected, this permutation fails entirely, serving primarily as a completeness check. It confirms that the improvement observed in the main analysis depends on the structural relationship between $\tilde{U}$ and treatment allocation.

All permutation experiments fail to reproduce the original improvement. This strongly suggests that the method does not operate through arbitrary or mechanistic adjustment. In particular, the $Y$-permutation experiment directly addresses the potential $Y \rightarrow U \rightarrow T \rightarrow Y$ feedback concern: because permuting $Y$ breaks the individual-level relationship while preserving arm-level survival distributions, the failure of the method under this manipulation demonstrates that incorporating $Y$ in the construction of $\tilde{U}$ does not automatically induce artificial correction. The improvement arises only when the genuine structure linking $X$, $Y$, and $T$ is preserved.

\subsection{Ablation Experiments}

To further understand which components of the $\tilde{U}$ construction drive the results, we conducted several ablation studies.

\paragraph{$\tilde{U} =$ signed outcome}

In this simplified variant, we replaced $\tilde{U}$ with the sign of the outcome only, testing whether directional information alone (risk versus protection) is sufficient to drive the correction.

\paragraph{$\tilde{U} =$ signed RMST}

Here, we used the continuous signed RMST without neighbor-based differencing. This preserves outcome magnitude but removes the comparative (neighbor-based) component of the construction.

\paragraph{Random neighbors (direction preserved)}

Finally, we computed $\tilde{U} $ using randomly selected neighbors rather than nearest neighbors in $X$-space, while preserving directionality (i.e., selecting frailer neighbors for anchors with no events and vice versa). In this case, the covariates $X$ no longer influence neighbor selection.

The sign-based experiments produce unstable and often problematic results. In particular, in GIST and BRT, these ablations substantially distort the HR estimates. They are also inconsistent in the CRLM multi-center analysis. This indicates that directional information alone is insufficient; the magnitude and structure of the neighbor comparison are essential.

Interestingly, the random-neighbor experiment yields results comparable to the full method in some datasets (e.g., CRLM), but performs substantially worse in others, particularly GIST and BRT. We interpret this heterogeneity as follows.

In datasets where the observed covariates $X$ have limited prognostic strength, the signal available for constructing meaningful neighborhoods is inherently weak. As a result, both nearest-neighbor and random direction-consistent constructions rely on similarly noisy information, leading to comparable performance. In contrast, in datasets such as GIST, where the covariates are strongly prognostic, accurately identifying nearest neighbors in $X$-space becomes essential. In these settings, replacing structured neighborhoods with random ones leads to a clear degradation in performance.


Taken together, the permutation and ablation studies demonstrate that:

\begin{itemize}
    \item The method does not operate via mechanical outcome injection.
    \item Individual-level alignment between $X$, $Y$, and $U$ is necessary.
    \item The neighbor-based construction contributes meaningfully beyond simple outcome transformations.
    \item The effectiveness of nearest-neighbor selection depends on the prognostic strength of the observed covariates.
\end{itemize}

These findings reinforce the interpretation of $U$ as a data-derived proxy for latent, pre-treatment summary of unobserved confounders, rather than an artifact of post-treatment outcome manipulation.

\subsection{Statistical Results}

\begin{table}[ht]
\centering
\caption{Comparison of balancing methods across datasets. $\Delta$ represents directional improvement in closeness to the benchmark RCT hazard ratio after adding $\tilde{U}$. Mean difference is reported on the log hazard ratio scale.}
\label{tab:method_comparison}
\begin{tabular}{llcc}
\hline
\textbf{Dataset} & \textbf{Method} & \textbf{Mean $\Delta$} & \textbf{Std. Error} \\ 
\hline
\multirow{3}{*}{GIST} 
& Matching & 0.334 & 0.012 \\ 
& Entropy Balancing & 0.256 & 0.022 \\
& IPTW & 0.326 & 0.026 \\
\hline
\multirow{3}{*}{RPS} 
& Matching & 0.084 & 0.008 \\ 
& Entropy Balancing & 0.592 & 0.031 \\
& IPTW & 0.903 & 0.007 \\
\hline
\multirow{3}{*}{CRLM} 
& Matching & 0.278 & 0.006 \\
& Entropy Balancing & 0.162 & 0.007 \\
& IPTW & 0.162 & 0.004 \\ 
\hline
\end{tabular}
\end{table}

\begin{table}[H]
\centering
\caption{Equivalence testing of treatment effect differences across datasets. Mean difference is calculated as $\Delta_{\text{shift}} = \log(HR^{\mathrm{aug}})-\log(HR^{\mathrm{RCT}})$.}
\label{tab:equivalence_loghr}
\begin{tabular}{llccc}
\toprule
\textbf{Dataset} & \textbf{Method} & \textbf{Mean Diff. (log HR)} & \textbf{Std. Error} & \textbf{95\% CI} \\
\midrule
\multirow{3}{*}{BRT}
& Matching & 0.075 & 0.0085 & [0.058, 0.092] \\
& Entropy Balancing & 0.183 & 0.0221 & [0.139, 0.227] \\
& IPTW & 0.035 & 0.0163 & [0.002, 0.067] \\
\midrule
\multirow{3}{*}{STRASS}
& Matching & -0.091 & 0.0058 & [-0.1025, -0.796] \\
& Entropy Balancing & -0.038 & 0.0095 & [-0.057, -0.019] \\
& IPTW & -0.066 & 0.0079 & [-0.081, -0.050] \\
\bottomrule
\end{tabular}
\end{table}

\begin{table}[H]
\centering
\caption{Comparison of balancing methods across datasets. $\Delta$ represents the reduction in mean pairwise absolute deviation after adjusting for $X, \tilde{U}$ for 6 CRLM centers. Mean difference is reported on the log hazard ratio scale.}
\label{tab:event_method_comparison}
\begin{tabular}{llcc}
\toprule
\textbf{Event} & \textbf{Method} & \textbf{Mean $\Delta$} & \textbf{Std. Error} \\
\midrule
\multirow{3}{*}{Mortality}
& Matching & 0.129 & 0.0087 \\
& Entropy Balancing & 0.098 & 0.0327 \\
& IPTW & 0.224 & 0.0029 \\
\midrule
\multirow{3}{*}{Recurrence}
& Matching & 0.139 & 0.0047 \\
& Entropy Balancing & 0.169 & 0.0360 \\
& IPTW & 0.121 & 0.0035 \\
\bottomrule
\end{tabular}
\end{table}

\end{document}